\documentclass[aps,prl,reprint,superscriptaddress]{revtex4-2}

\usepackage{amsmath}
\usepackage{mathtools}
\usepackage{amssymb}
\usepackage{graphicx}
\usepackage{bm}
\usepackage[colorlinks,linkcolor=blue,citecolor=blue,urlcolor=blue,breaklinks=true]{hyperref}
\usepackage{booktabs}
\usepackage{multirow}
\usepackage[dvipsnames]{xcolor}
\usepackage{ragged2e}
\usepackage{orcidlink}
\usepackage[T1]{fontenc}
\usepackage{mathrsfs}
\usepackage{graphicx}

\newcommand{\Tr}{\operatorname{Tr}}
\newcommand{\kB}{k_{\rm B}}
\newcommand{\cD}{\mathcal D}
\newcommand{\figpanel}[2]{\hyperref[#1]{\ref*{#1}(#2)}}

\begin{document}

\title{Physical-Work Fluctuation Relations from Accessible Quantum Macrostates}

\author{Borhan Ahmadi\orcidlink{0000-0002-2787-9321}}
\email{borhan.ahmadi@ug.edu.pl}
\affiliation{International Centre for Theory of Quantum Technologies,
University of Gda\'nsk, ul.~prof. Marii Janion 4,
80-309 Gda\'nsk, Poland}

\date{\today}

\begin{abstract}
Jarzynski's equality recovers an equilibrium free-energy difference from nonequilibrium work trajectories, but its exponential average can converge very slowly because rare trajectories carry large weight. We show that coarse thermodynamic information measured at the nonequilibrium endpoint can reduce this sampling burden while keeping the same microscopic trajectories and the same free-energy target. The endpoint mean energy and a coarse spatial record define a maximum-entropy state and select one member of an exact family of fluctuation relations. That physically selected member provides an exact statistical control for the ordinary Jarzynski estimator, even when the final energy and retained record do not commute. In a finite Bose--Hubbard system, this control substantially lowers the finite-confidence sampling cost, while a small controlled departure from exact endpoint matching produces a square-root information--sampling tradeoff.
\end{abstract}

\maketitle

\textit{Introduction.---}
Thermodynamics is useful because it lets us describe a system with a few accessible quantities instead of its full microscopic state.
Statistical mechanics makes this reduction precise by relating macroscopic data to the microscopic states that are compatible with them.
Jaynes gave a simple information-theoretic rule for this situation: when only a set of constraints is known, the maximum-entropy state keeps that information without adding assumptions that are not supported by the data~\cite{Jaynes1957}.
Projection methods give a related dynamical picture, in which the exact state is separated into a part fixed by chosen observables and a remaining unresolved part~\cite{Zwanzig1960,Robertson1966}.
For a small quantum system this distinction is operational: the microscopic evolution can remain fully quantum even when the thermodynamic description retains only a few measured observables.
Related quantum approaches have built thermodynamic descriptions from coarse or accessible information without requiring full state tomography~\cite{Safranek2019,StrasbergWinter2021,Buscemi2023,Meier2025,Schindler2025,Xiao2026}.

A different reduction appears in nonequilibrium work relations.
Jarzynski showed that work values collected during a driven process can determine an equilibrium free-energy difference \(\Delta F\), and Crooks related the same structure to forward and reversed processes~\cite{Jarzynski1997,Crooks1999}.
In quantum systems, the standard two-point energy measurement (TPM) assigns a work value to each run from its initial and final energy outcomes~\cite{Talkner2007,Campisi2011}.
The difficulty is not the exactness of the relation but its sampling.
The exponential average can be controlled by rare trajectories that occur with very small probability, so a reliable free-energy estimate may require a very large number of runs~\cite{Gore2003,Jarzynski2006}.

A broad literature has developed around this sampling problem.
The cost can be reduced by changing the driving protocol, introducing escorted dynamics, choosing better intermediate states, or optimizing the Hamiltonian path~\cite{VaikuntanathanJarzynski2008,Shenfeld2009,BeyerGrubmuller2026}.
Learned transport maps provide another route to low-variance free-energy estimators~\cite{DuEtAl2025FEAT}.
Other approaches leave already generated trajectories in place and improve how they are analyzed.
These include optimal path-ensemble estimators~\cite{MinhChodera2009}, protocol post-processing based on nonequilibrium densities~\cite{Minh2009,MinhVaikuntanathan2011}, conditioning on measurement outcomes~\cite{AsbanRahav2017}, and conditioning on trajectory classes~\cite{Wimsatt2025}.
Most recently, Lee and Jarzynski constructed a family of virtually escorted work-like quantities that use the same underlying trajectories and recover the same free-energy difference, with the auxiliary construction chosen to improve convergence~\cite{LeeJarzynski2026}.
Thus, changing the statistical analysis of fixed trajectories is itself an established strategy.

Quantum fluctuation relations have meanwhile been extended in several other directions.
General formulations apply to quantum channels and to more general dynamical references~\cite{KwonKim2019,CuginiGuarnieri2026}.
Coarse-grained work measurements and thermodynamics based directly on restricted sets of accessible observables have also been developed~\cite{Rubino2026,RignonBretElouard2026}.
Recent fluctuation relations further show that noncommuting thermodynamic quantities can produce genuinely quantum corrections~\cite{ScandiManzano2026,RodriguesLutz2026}.
These results broaden both the allowed reference states and the observables that may enter a fluctuation relation.
They leave open, however, a different question: can a thermodynamic state that is independently fixed by a few measurements of the actual nonequilibrium endpoint be turned into a statistical resource for estimating the ordinary Jarzynski free-energy difference?

We address this question without changing the physical work protocol or the microscopic trajectories. After the drive, the measured endpoint energy and coarse spatial record define a maximum-entropy representative of the undisturbed nonequilibrium state~\cite{Ahmadi2026Endogenous,Ahmadi2026FiniteParticle}. This physical reference leads to an exact work-and-record fluctuation relation while keeping the ordinary fine-grained TPM work unchanged.
The endpoint-matching relation belongs to a larger exact family that also contains Jarzynski's relation. Its members can be used as control variates for the same free-energy estimate~\cite{NguyenMinh2020}, but the thermodynamic member is singled out by the measured endpoint rather than by statistical convenience. In the finite Bose--Hubbard system studied below, this endpoint information substantially lowers the finite-confidence sampling cost, and allowing a small controlled departure from it lowers the cost further. The resulting information--sampling frontier has a square-root onset near exact endpoint matching, revealing how a small loss of thermodynamic fidelity can produce a much larger statistical gain.
\begin{figure*}[t]
\centering
\includegraphics[width=\textwidth]{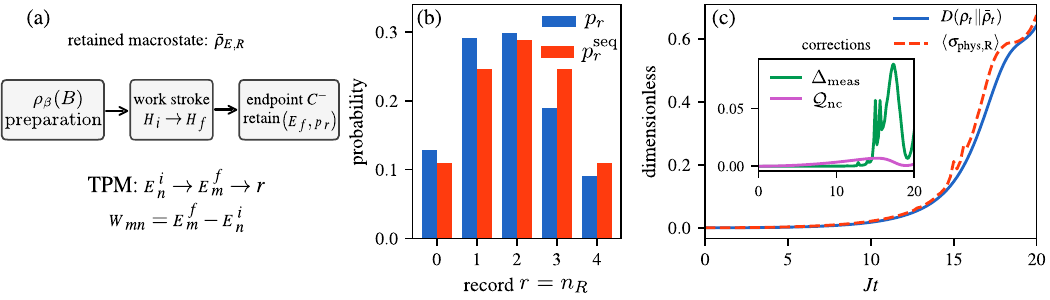}
\caption{
Physical work and the accessible endpoint record.
(a) The system starts from the canonical state \(B\), is thermally isolated, and is driven from \(H_i\) to \(H_f\), reaching the nonequilibrium endpoint \(C^-\). Undisturbed copies determine the endpoint mean energy and spatial record, which define \(\bar\rho_{E,R}\). Separate fluctuation runs measure \(E_n^i\to E_m^f\to r\) and assign the ordinary TPM work \(W_{mn}=E_m^f-E_n^i\).
(b) The undisturbed record \(p_r\) differs from the record \(p_r^{\rm seq}\) seen after the final energy measurement. This difference is the measurement-order effect that enters the mean fluctuation relation.
(c) Along independently prepared truncated drives, the unresolved information \(D(\rho_t\Vert\bar\rho_t)\) grows together with \(\langle\sigma_{\rm phys,R}\rangle\). The inset separates the smaller measurement-order and noncommutativity corrections.
}
\label{fig:physical-work-record}
\end{figure*}

\textit{Accessible quantum macrostate.---}
We consider a finite isolated quantum system with exact state \(\rho\). The thermodynamic observer does not retain the full matrix \(\rho\). Instead, the observer keeps the mean energy \(E=\Tr(H\rho)\) and the probabilities \(p_r=\Tr(P_r\rho)\) of a coarse projective record \(R\), with \(\sum_rP_r=I\). These measured quantities define the maximum-entropy state compatible with the retained information,
\begin{equation}
\bar\rho_{E,R}
=
\frac{e^{-\beta^\star H-A}}{Z},
\quad
A
=
\sum_r\alpha_rP_r.
\label{eq:maxent-main}
\end{equation}
Here \(Z=\Tr e^{-\beta^\star H-A}\). The multipliers \(\beta^\star\) and \(\alpha_r\) are chosen so that \(\bar\rho_{E,R}\) reproduces the measured \(E\) and \(p_r\). Thus \(\bar\rho_{E,R}\) contains exactly the thermodynamic information we keep and no additional microscopic detail. Away from equilibrium, \(\beta^\star\) is only a maximum-entropy multiplier and should not be read as a physical inverse temperature.

The exact state and its thermodynamic representative need not be the same. Their relative entropy \(D(\rho\Vert\bar\rho_{E,R})\) measures the microscopic information left unresolved by the retained record. In the language of Ref.~\cite{Ahmadi2026Endogenous}, this missing part is carried by an unresolved operator. Here we write
\(
\chi_{E,R}(t) = \rho_t-\bar\rho_{E,R}(t).
\)
At the same time, \(\chi_{E,R}(t)\) has zero contribution to the retained energy and record. It can nevertheless affect later accessible observables after further unitary evolution (see Note~\ref{supp:note-targetB-derivation} of the SM \cite{SuppMat} for the exact link to that dynamical return).

We next connect this endpoint description to ordinary microscopic work trajectories. The initial state is canonical,
\(
\rho_i
= e^{-\beta_iH_i}/Z_i,
\)
with
\(
Z_i
=
\Tr e^{-\beta_iH_i}.
\)
An isolated control changes \(H_i\) to \(H_f\) through a unitary \(U\), so the undisturbed endpoint is \(\rho_f=U\rho_iU^\dagger\). The mean work is already fixed by the mean-energy change and does not require TPM. We use TPM for a different reason: it gives a work value for each realization. If the energy outcomes are \(E_n^i\) and \(E_m^f\), then \(W_{mn} = E_m^f-E_n^i.\)
Because the initial state is canonical, the first energy measurement does not change the ensemble. After the final energy measurement, we read the record \(r\). One fluctuation run therefore resolves \(E_n^i\to E_m^f\to r\), as shown in Fig.~\figpanel{fig:physical-work-record}{a}.

The endpoint thermodynamic state must be fixed before this fluctuation measurement is allowed to disturb it. We therefore use undisturbed endpoint copies to determine the final mean energy and record, fit the corresponding multipliers, and then freeze those parameters. Separate copies are used for the TPM-and-record runs. This separation is essential when the final energy and record do not commute.

\textit{Physical-work-and-record fluctuation relation.---}
At the endpoint, the retained data define
\begin{equation}
\bar\rho_f
=
\frac{e^{-\beta_f^\star H_f-A_f}}{Z_f},
\quad
A_f
=
\sum_r\alpha_{r,f}P_r.
\label{eq:endpoint-representative-main}
\end{equation}
Here \(Z_f=\Tr e^{-\beta_f^\star H_f-A_f}\). The final energy measurement dephases the state before the record is read. The record sampled by the sequential protocol is therefore
\(p_{r,f}^{\rm seq}=\Tr[P_r\cD_{H_f}(\rho_f)]\), which can differ from the undisturbed thermodynamic record \(p_{r,f}=\Tr(P_r\rho_f)\).

The same measurement order also changes the normalization entering the trajectory average,
\begin{equation}
Z_f^{\rm seq}
=
\Tr
\left[
e^{-\beta_f^\star H_f}
e^{-A_f}
\right],
\quad
\mathcal Q_{\rm nc}
=
\ln\frac{Z_f^{\rm seq}}{Z_f}
\geq
0.
\label{eq:qnc-main}
\end{equation}
The quantity \(\mathcal Q_{\rm nc}\) is therefore not an additional thermodynamic irreversibility. It measures the difference between the joint exponential defining the thermodynamic representative and the ordered product generated by measuring energy before the record. It vanishes when \(H_f\) and \(A_f\) commute.

Define \(\Psi_i=-\ln Z_i\), \(\Psi_f=-\ln Z_f\), and \(\Delta\Psi=\Psi_f-\Psi_i\). For one trajectory \((n,m,r)\), let
\begin{equation}
\sigma_{\rm phys,R}
=
\beta_iW_{mn}
+
\left(
\beta_f^\star-\beta_i
\right)
E_m^f
+
\alpha_{r,f}
+
\mathcal Q_{\rm nc}
-
\Delta\Psi.
\label{eq:sigma-main}
\end{equation}
Its exponential average is exactly normalized,
\begin{equation}
\left\langle
e^{-\sigma_{\rm phys,R}}
\right\rangle
=
1.
\label{eq:ift-main}
\end{equation}
The cancellation begins in the same way as in Jarzynski's equality: the initial-energy factor removes the canonical preparation weight. Completeness of the energy projectors and unitarity then eliminate the detailed propagator, leaving precisely the sequential endpoint normalization in Eq.~\eqref{eq:qnc-main}. The full derivation, including degeneracies and the gauge freedom of the record multipliers, is given in Note~\ref{supp:note-targetB-derivation} of the SM \cite{SuppMat}.

The mean of the same trajectory variable separates three distinct effects,
\begin{equation}
\left\langle
\sigma_{\rm phys,R}
\right\rangle
=
D
\left(
\rho_f
\Vert
\bar\rho_f
\right)
+
\Delta_{\rm meas}
+
\mathcal Q_{\rm nc},
\label{eq:mean-decomposition-main}
\end{equation}
where
\(\Delta_{\rm meas}=\sum_r\alpha_{r,f}(p_{r,f}^{\rm seq}-p_{r,f})\).
The first term is information already unresolved by the coarse thermodynamic description before the fluctuation measurement is performed. The second measures how the final energy measurement changes the weighted record, while the third measures the corresponding change in normalization. If the record projectors commute with \(H_f\), the energy measurement leaves the record unchanged and both measurement-sequence corrections vanish. The decomposition therefore separates pre-existing unresolved information from effects introduced by the order of the endpoint measurements. Neither \(\sigma_{\rm phys,R}\) nor its full mean is, in general, identified with physical entropy generation.

We now give a concrete physical realization of this separation in a finite interacting quantum system. In this example, the unresolved-information term can be identified independently with the thermodynamic irreversibility of the isolated work stroke, while the two remaining terms can be seen directly as consequences of the noncommuting endpoint measurements.
\begin{figure*}[t] \centering \includegraphics[width=\textwidth]{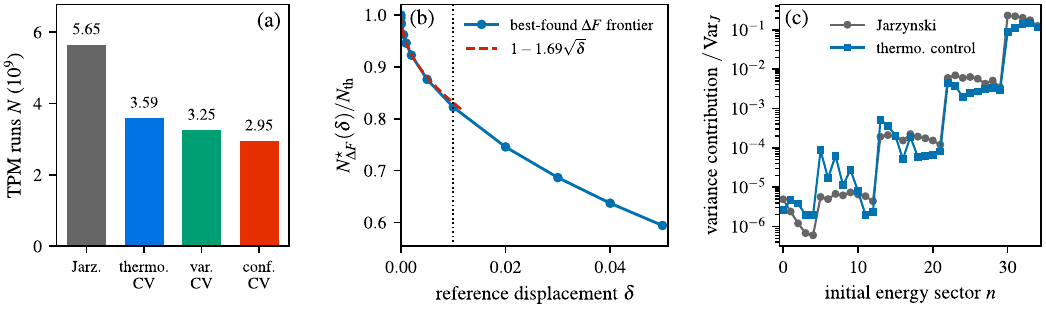} \caption{ Endpoint information used as an exact control for the same Jarzynski free-energy target. (a) Model-conditioned TPM repetition benchmarks for the same free-energy difference. Direct Jarzynski sampling is compared with the endpoint-matching thermodynamic control and with variance- and finite-confidence-optimized controls at the same displacement budget. Lower bars mean fewer production trajectories. (b) Best-found finite-confidence cost for the same free-energy target as the exact control reference moves a relative-entropy distance \(\delta\) away from the measured thermodynamic endpoint. The dashed curve is the local square-root law. (c) Contribution of each initial energy sector to the estimator variance, normalized by the total direct-Jarzynski variance. The thermodynamic control reduces the large rare-sector contributions without changing the sampled trajectories. } \label{fig:information-sampling} \end{figure*} 

\textit{Finite Bose--Hubbard working medium.---}
We consider the isolated \(B\rightarrow C^-\) work stroke of Ref.~\cite{Ahmadi2026FiniteParticle} as a self-contained test of the relation. The working medium contains \(Q=4\) bosons on an \(L=4\) open chain,
\(
H(\Lambda)
=
-J
\sum_{j=1}^{L-1}
\left(
a_j^\dagger a_{j+1}
+
a_{j+1}^\dagger a_j
\right)
+
\frac{U}{2}
\sum_{j=1}^{L}
n_j(n_j-1)
+
\Lambda N_R,
\)
with \(N_R=n_3+n_4\). At \(B\), the system is canonical at \(\beta_iJ=0.2\) with \(\Lambda_0=20J\), which strongly biases the particles to the left. It is then thermally isolated while \(\Lambda\) is lowered linearly to zero over \(J\tau=20\), with \(U/J=1.3/3\). No heat crosses the boundary during the stroke; the change in mean energy is work transferred through the external control.

The drive produces the nonequilibrium endpoint
\(\rho_f=\rho_{C^-}=U_{BC}\rho_BU_{BC}^\dagger\) of \(H(0)\). The retained record is \(r=n_R\in\{0,1,2,3,4\}\). Although the fixed-particle Hilbert space has dimension \(35\), the thermodynamic description keeps only the mean energy and four independent record probabilities. Because the initial canonical state already equals its energy-and-record representative and unitary evolution preserves the von Neumann entropy, the endpoint information gap is exactly the stroke irreversibility of Ref.~\cite{Ahmadi2026FiniteParticle},
\begin{equation}
D
\left(
\rho_{C^-}
\Vert
\bar\rho_{E,R}(C^-)
\right)
=
\frac{\Sigma_{BC}}{\kB},
\label{eq:D-Sigma-main}
\end{equation}
where \(k_{\rm B}\) is the Boltzmann constant. Thus the first term in Eq.~\eqref{eq:mean-decomposition-main} has a direct thermodynamic meaning in this model, while the remaining two terms arise from the fluctuation-measurement sequence.

Figure~\figpanel{fig:physical-work-record}{a} separates these roles operationally: undisturbed copies determine the thermodynamic endpoint, while independent copies provide the TPM-and-record trajectories. Figure~\figpanel{fig:physical-work-record}{b} then shows the measurement backaction directly. At the symmetric final Hamiltonian, energy dephasing removes the coherence supporting the left--right asymmetry of the undisturbed record, so the sequential distribution becomes symmetric about \(r=2\). Their difference is the physical origin of \(\Delta_{\rm meas}\).

Figure~\figpanel{fig:physical-work-record}{c} shows the same separation during the drive. The unresolved information starts from zero and grows as the unitary dynamics creates microscopic structure not fixed by the instantaneous energy and record, while the two measurement-sequence corrections remain smaller. At the endpoint,
\begin{equation}
\begin{aligned}
D(\rho_f\Vert\bar\rho_f)
&=
0.641610,
\quad
\Delta_{\rm meas}
=
0.031325,
\\
\mathcal Q_{\rm nc}
&=
0.001879,
\quad
\left\langle\sigma_{\rm phys,R}\right\rangle
=
0.674814.
\end{aligned}
\label{eq:BH-endpoint-main}
\end{equation}
The complete time-resolved fits, independent propagation checks, and endpoint parameters are given in Note~\ref{supp:note-BH-validation} of the SM \cite{SuppMat}. Having established the physical meaning of the endpoint-selected relation, we now ask whether it can also reduce the sampling cost of the ordinary Jarzynski free-energy estimate without changing the work protocol or the sampled trajectories.

\textit{Exact thermodynamic controls for the same free-energy target.---}
We now return to the sampling problem. The endpoint fit picks one physically meaningful fluctuation relation, but the same cancellation works for a larger family. For any fixed \(\bm\vartheta=(\beta,\bm\alpha)\), let \(A(\bm\alpha)=\sum_r\alpha_rP_r\) and \(Z_f^{\rm seq}(\bm\vartheta)=\Tr[e^{-\beta H_f}e^{-A(\bm\alpha)}]\). Define
\begin{equation}
\sigma_{\bm\vartheta}
=
\beta_iW_{mn}
+
(\beta-\beta_i)E_m^f
+
\alpha_r
+
\ln\frac{Z_f^{\rm seq}(\bm\vartheta)}{Z_i}.
\label{eq:family-sigma-main}
\end{equation}
Every fixed member obeys
\begin{equation}
\left\langle e^{-\sigma_{\bm\vartheta}}\right\rangle
=
1.
\label{eq:family-ift-main}
\end{equation}
The thermodynamic member \(\bm\vartheta_{\rm th}=(\beta_f^\star,\bm\alpha_f)\) is special because its maximum-entropy state reproduces the measured mean energy and coarse record of the undisturbed endpoint. Its parameters are therefore fixed by the physical endpoint itself, before any statistical optimization, rather than chosen to reduce the fluctuations of the Jarzynski estimator. Any sampling advantage it provides thus comes from a physically selected reference. The Jarzynski member \(\bm\vartheta_J=(\beta_i,\bm 0)\) instead gives \(\sigma_{\bm\vartheta_J}=\beta_i(W-\Delta F)\). All members use the same trajectory probabilities; only the exponential weight attached to each trajectory changes. Unlike coarse-grained TPM constructions, where limited resolution changes the measured work variable itself, here the TPM work remains fully resolved; coarse information enters only through the independently calibrated nonequilibrium endpoint.

The new step is to use these exact identities as controls for one common target. The ordinary Jarzynski weight and its mean are
\begin{equation}
X_J
=
e^{-\beta_iW},
\quad
R_J
\equiv
\left\langle X_J\right\rangle
=
e^{-\beta_i\Delta F}.
\label{eq:jarzynski-target-main}
\end{equation}
Because Eq.~\eqref{eq:family-ift-main} has unit mean, each family member gives a zero-mean random variable,
\begin{equation}
C_{\bm\vartheta}
=
e^{-\sigma_{\bm\vartheta}}-1,
\quad
\left\langle C_{\bm\vartheta}\right\rangle
=
0.
\label{eq:control-zero-main}
\end{equation}
A zero-mean control can change fluctuations without changing the desired mean. For any fixed coefficient \(c\), we therefore estimate the Jarzynski ratio with
\begin{equation}
\widehat R_J(\bm\vartheta,c)
=
\frac{1}{N}
\sum_{j=1}^{N}
\left[
e^{-\beta_iW_j}
-
c\left(e^{-\sigma_{\bm\vartheta,j}}-1\right)
\right].
\label{eq:control-estimator-main}
\end{equation}
Its expectation is still \(R_J\). The control helps only through correlations: when its fluctuations track part of the large Jarzynski fluctuations, subtraction reduces the spread and the tail cost of the estimator (see Note~\ref{supp:note-sampling-frontier} of the SM \cite{SuppMat} for the analytical analysis). The physical thermodynamic member is valuable because the endpoint measurement selects it independently of this statistical purpose. A useful control must also be experimentally accessible: variance reduction is of little practical value if evaluating the control costs more than the sampling it saves.

We next ask whether a nearby exact control can perform even better. The distance from the physical thermodynamic reference is measured by
\begin{equation}
\delta
=
D\!\left(
\bar\rho_{\rm th}
\Vert
\bar\rho_{\bm\vartheta}
\right),
\quad
\bar\rho_{\bm\vartheta}
=
\frac{e^{-\beta H_f-A(\bm\alpha)}}{Z_f(\bm\vartheta)}.
\label{eq:delta-main}
\end{equation}
Thus \(\delta=0\) means that the control reference exactly matches the measured thermodynamic endpoint. Increasing \(\delta\) allows the control to give up a controlled amount of endpoint fidelity in exchange for a possible sampling gain.

For each allowed \(\delta\), we minimize a two-sided Cram\'er--Chernoff repetition benchmark for estimating \(R_J\) to fixed relative accuracy and confidence. We optimize both the control reference and the coefficient \(c\). These are model-conditioned production-run benchmarks after independent endpoint calibration; the raw estimator in Eq.~\eqref{eq:control-estimator-main} does not require prior knowledge of \(\Delta F\). In an experiment, \(c\) must be fixed from an independent pilot or calibration sample before the production runs (see Note~\ref{supp:note-sampling-frontier} of the SM \cite{SuppMat} for the full construction and resource accounting).

Figure~\figpanel{fig:information-sampling}{a} compares four ways of estimating the same free-energy difference. Direct Jarzynski sampling has the largest finite-confidence cost. Using the endpoint-matching thermodynamic member as a control lowers the benchmark from \(5.6463\times10^9\) to \(3.5873\times10^9\) runs, without moving away from the measured endpoint. At the common displacement budget \(\delta=0.01\), variance optimization gives \(3.2500\times10^9\) runs, while direct optimization of the full two-sided finite-confidence cost gives \(2.9496\times10^9\). The latter is a \(47.76\%\) reduction relative to direct Jarzynski sampling. The difference between the last two bars shows why a tail-sensitive objective cannot be replaced by variance alone.

Figure~\figpanel{fig:information-sampling}{b} shows the common-target information--sampling frontier. Close to the measured thermodynamic member (see Note~\ref{supp:note-sampling-frontier} of the SM \cite{SuppMat}),
\begin{equation}
1-
\frac{N_{\Delta F}^{\star}(\delta)}{N_{\rm th}}
=
C_{\Delta F}\sqrt{\delta}
+
O(\delta),
\quad
C_{\Delta F}\simeq1.69.
\label{eq:sqrt-law-main}
\end{equation}
The square-root onset has a simple origin.
Near the thermodynamic member, quantum relative entropy is locally quadratic and induces the Bogoliubov--Kubo--Mori information metric~\cite{Petz1994,Abiuso2020}, while the sampling cost can change already at first order. A parameter shift of order \(\sqrt\delta\) therefore costs only order \(\delta\) in endpoint fidelity but changes the optimized sampling cost at order \(\sqrt\delta\). At \(\delta=0.01\), the optimized control lowers the cost by \(17.78\%\) relative to exact thermodynamic matching, while the estimated quantity remains the same \(\Delta F\).

Figure~\figpanel{fig:information-sampling}{c} shows where the gain comes from microscopically. Direct Jarzynski sampling is strongly affected by rare high-energy initial sectors. The thermodynamic control is correlated with those large work weights and subtracts part of their fluctuation on the same runs. The rare-sector contribution to the variance is therefore reduced without discarding trajectories, changing the protocol, or changing the target free energy.

The canonical initial state makes the construction especially clean because the first energy measurement does not change it. For a general initial state, the first TPM measurement produces the dephased state \(\rho_i^{\rm d}=\cD_{H_i}[\rho_i]\). The exact relation then needs an additional energy-resolved correction \(\Gamma_i=\ln\rho_i^{\rm d}-\ln\tau_i^{(0)}\), where \(\tau_i^{(0)}=e^{-\beta_0H_i}/Z_i^{(0)}\) is an auxiliary Gibbs reference. The retained initial record must contain enough information to determine this correction. For the unrestricted \(35\)-dimensional Bose--Hubbard sector, the natural \((E,R)\) record leaves a \(29\)-dimensional family of compatible energy populations and is therefore not sufficient. The arbitrary-state theorem, an explicit same-\((E,R)\) counterexample, the rank-deficient support correction, and the initial calibration cost are given in Note~\ref{supp:note-arbitrary-initial} of the SM \cite{SuppMat}.

\textit{Discussion and summary.---}
We have kept the microscopic and thermodynamic descriptions separate. The work values remain the ordinary fine-grained TPM values, while the endpoint mean energy and coarse record define a maximum-entropy thermodynamic state. When the final energy and record do not commute, the fluctuation relation shows exactly how unresolved endpoint information, measurement disturbance, and noncommuting normalization enter the same trajectory average.
The main statistical result is that the endpoint-matching relation can do more than provide another exact normalization. Because its exponential weight has a known mean, it becomes an exact control for the ordinary Jarzynski estimator. The physical endpoint selects this control independently, and nearby exact controls provide a controlled trade between endpoint fidelity and free-energy sampling cost. All points on this frontier use the same trajectories and estimate the same equilibrium free-energy difference.
The reported repetition counts isolate the production TPM cost after endpoint calibration and after the control coefficient has been fixed. A full experiment must also count those calibration and pilot copies. Within that scope, the Bose--Hubbard example shows that a small amount of accessible nonequilibrium information can substantially reduce the rare-event burden of the standard Jarzynski task without changing the microscopic dynamics, the work definition, or the free-energy target.

\textit{Acknowledgments.---}
BA acknowledges support from IRA Program (project no. FENG.02.01-IP.05-0006/23) financed by the FENG program 2021-2027, Priority FENG.02, Measure FENG.02.01., with the support of the FNP.

\textit{Data and code availability.---}
The numerical code used to generate the data and figures in this work is available at \url{https://github.com/Borhan19/PWFluctuations/releases/tag/v1.0-submission}. 
The repository contains the simulation scripts, plotting routines, and parameter files required to reproduce the results presented in the main text and Supplemental Material.

\bibliography{References}

@misc{SuppMat,
  note = {See Supplemental Material for the full derivations, numerical validation, information--sampling analysis, and arbitrary-initial-state extension.}
}

@article{Jarzynski1997,
  author  = {Jarzynski, Christopher},
  title   = {Nonequilibrium Equality for Free Energy Differences},
  journal = {Physical Review Letters},
  volume  = {78},
  number  = {14},
  pages   = {2690--2693},
  year    = {1997},
  doi     = {10.1103/PhysRevLett.78.2690}
}

@article{Ahmadi2026Endogenous,
  author        = {Borhan Ahmadi},
  title         = {Thermodynamic Irreversibility from Inaccessible Endogenous Quantum Histories},
  journal        = {arXiv: 2609.03748},
  year           = {2026},
  url           = {https://doi.org/10.48550/arXiv.2609.03748}
}

@article{Ahmadi2026FiniteParticle,
  author        = {Borhan Ahmadi},
  title         = {Finite-Particle Quantum Reduction of Thermodynamic Irreversibility},
  journal        = {arXiv: 2609.10211},
  year           = {2026},
  url           = {https://doi.org/10.48550/arXiv.2609.10211}
}

@article{Jaynes1957,
  author  = {Jaynes, E. T.},
  title   = {Information Theory and Statistical Mechanics},
  journal = {Physical Review},
  volume  = {106},
  number  = {4},
  pages   = {620--630},
  year    = {1957},
  doi     = {10.1103/PhysRev.106.620},
  url     = {https://doi.org/10.1103/PhysRev.106.620}
}

@article{Zwanzig1960,
  author  = {Zwanzig, Robert},
  title   = {Ensemble Method in the Theory of Irreversibility},
  journal = {The Journal of Chemical Physics},
  volume  = {33},
  number  = {5},
  pages   = {1338--1341},
  year    = {1960},
  doi     = {10.1063/1.1731409},
  url     = {https://doi.org/10.1063/1.1731409}
}

@article{Robertson1966,
  author  = {Robertson, Baldwin},
  title   = {Equations of Motion in Nonequilibrium Statistical Mechanics},
  journal = {Physical Review},
  volume  = {144},
  number  = {1},
  pages   = {151--161},
  year    = {1966},
  doi     = {10.1103/PhysRev.144.151},
  url     = {https://doi.org/10.1103/PhysRev.144.151}
}

@article{Safranek2019,
  author  = {{\v S}afr{\'a}nek, Dominik and Deutsch, J. M. and Aguirre, Anthony},
  title   = {Quantum Coarse-Grained Entropy and Thermodynamics},
  journal = {Physical Review A},
  volume  = {99},
  number  = {1},
  pages   = {010101},
  year    = {2019},
  doi     = {10.1103/PhysRevA.99.010101},
  url     = {https://doi.org/10.1103/PhysRevA.99.010101}
}

@article{StrasbergWinter2021,
  author  = {Strasberg, Philipp and Winter, Andreas},
  title   = {First and Second Law of Quantum Thermodynamics: A Consistent Derivation Based on a Microscopic Definition of Entropy},
  journal = {PRX Quantum},
  volume  = {2},
  number  = {3},
  pages   = {030202},
  year    = {2021},
  doi     = {10.1103/PRXQuantum.2.030202},
  url     = {https://doi.org/10.1103/PRXQuantum.2.030202}
}

@article{Buscemi2023,
  author  = {Buscemi, Francesco and Schindler, Joseph and {\v S}afr{\'a}nek, Dominik},
  title   = {Observational Entropy, Coarse-Grained States, and the Petz Recovery Map: Information-Theoretic Properties and Bounds},
  journal = {New Journal of Physics},
  volume  = {25},
  number  = {5},
  pages   = {053002},
  year    = {2023},
  doi     = {10.1088/1367-2630/accd11},
  url     = {https://doi.org/10.1088/1367-2630/accd11}
}

@article{Meier2025,
  author  = {Meier, Florian and Rivlin, Tom and Debarba, Tiago and Xuereb, Jake and Huber, Marcus and Lock, Maximilian P. E.},
  title   = {Emergence of a Second Law of Thermodynamics in Isolated Quantum Systems},
  journal = {PRX Quantum},
  volume  = {6},
  number  = {1},
  pages   = {010309},
  year    = {2025},
  doi     = {10.1103/PRXQuantum.6.010309},
  url     = {https://doi.org/10.1103/PRXQuantum.6.010309}
}

@article{Schindler2025,
  author        = {Schindler, Joseph and Strasberg, Philipp and Galke, Niklas and Winter, Andreas and Jabbour, Michael G.},
  title         = {Unification of Observational Entropy with Maximum Entropy Principles},
  year          = {2025},
  journal        = {arXiv: 2503.15612},
  url           = {https://arxiv.org/abs/2503.15612}
}

@article{Rubino2026,
  author  = {Rubino, Giulia and Brukner, {{\v C}aslav} and Manzano, Gonzalo},
  title   = {Coarse-Grained Quantum Thermodynamics: Observation-Dependent Quantities, Observation-Independent Laws},
  journal = {Physical Review Research},
  volume  = {8},
  number  = {2},
  pages   = {023284},
  year    = {2026},
  doi     = {10.1103/gcyn-18tv},
  url     = {https://doi.org/10.1103/gcyn-18tv}
}

@article{RignonBretElouard2026,
  author        = {Rignon-Bret, Antoine and Elouard, Cyril},
  title         = {Quantum Stochastic Thermodynamics of Macroscopic Systems: An Algebraic Approach},
  year          = {2026},
  journal        = {arXiv: 2607.09242},
  url           = {https://arxiv.org/abs/2607.09242}
}

@article{Crooks1999,
  author  = {Crooks, Gavin E.},
  title   = {Entropy Production Fluctuation Theorem and the Nonequilibrium Work Relation for Free Energy Differences},
  journal = {Physical Review E},
  volume  = {60},
  number  = {3},
  pages   = {2721--2726},
  year    = {1999},
  doi     = {10.1103/PhysRevE.60.2721},
  url     = {https://doi.org/10.1103/PhysRevE.60.2721}
}

@article{Talkner2007,
  author  = {Talkner, Peter and Lutz, Eric and H{\"a}nggi, Peter},
  title   = {Fluctuation Theorems: Work Is Not an Observable},
  journal = {Physical Review E},
  volume  = {75},
  number  = {5},
  pages   = {050102},
  year    = {2007},
  doi     = {10.1103/PhysRevE.75.050102},
  url     = {https://doi.org/10.1103/PhysRevE.75.050102}
}

@article{Campisi2011,
  author  = {Campisi, Michele and H{\"a}nggi, Peter and Talkner, Peter},
  title   = {Colloquium: Quantum Fluctuation Relations: Foundations and Applications},
  journal = {Reviews of Modern Physics},
  volume  = {83},
  number  = {3},
  pages   = {771--791},
  year    = {2011},
  doi     = {10.1103/RevModPhys.83.771},
  url     = {https://doi.org/10.1103/RevModPhys.83.771}
}

@article{Gore2003,
  author  = {Gore, Jeff and Ritort, Felix and Bustamante, Carlos},
  title   = {Bias and Error in Estimates of Equilibrium Free-Energy Differences from Nonequilibrium Measurements},
  journal = {Proceedings of the National Academy of Sciences},
  volume  = {100},
  number  = {22},
  pages   = {12564--12569},
  year    = {2003},
  doi     = {10.1073/pnas.1635159100},
  url     = {https://doi.org/10.1073/pnas.1635159100}
}

@article{Jarzynski2006,
  author  = {Jarzynski, Christopher},
  title   = {Rare Events and the Convergence of Exponentially Averaged Work Values},
  journal = {Physical Review E},
  volume  = {73},
  number  = {4},
  pages   = {046105},
  year    = {2006},
  doi     = {10.1103/PhysRevE.73.046105},
  url     = {https://doi.org/10.1103/PhysRevE.73.046105}
}

@article{Xiao2026,
  author  = {Xiao, Yuanyuan and Jiang, Jian-Hua and Liu, Junjie},
  title   = {Quantum information-cost relations and fluctuations beyond thermal environments: A thermodynamic inference approach},
  journal = {Physical Review A},
  volume  = {113},
  pages   = {032214},
  year    = {2026},
  doi     = {10.1103/qrvg-nydb},
  url     = {https://doi.org/10.1103/qrvg-nydb}
}

@article{VaikuntanathanJarzynski2008,
  author  = {Vaikuntanathan, Suriyanarayanan and Jarzynski, Christopher},
  title   = {Escorted Free Energy Simulations: Improving Convergence by Reducing Dissipation},
  journal = {Physical Review Letters},
  volume  = {100},
  pages   = {190601},
  year    = {2008},
  doi     = {10.1103/PhysRevLett.100.190601},
  url     = {https://doi.org/10.1103/PhysRevLett.100.190601}
}

@article{Shenfeld2009,
  author  = {Shenfeld, Daniel K. and Xu, Huafeng and Eastwood, Michael P. and Dror, Ron O. and Shaw, David E.},
  title   = {Minimizing thermodynamic length to select intermediate states for free-energy calculations and replica-exchange simulations},
  journal = {Physical Review E},
  volume  = {80},
  pages   = {046705},
  year    = {2009},
  doi     = {10.1103/PhysRevE.80.046705},
  url     = {https://doi.org/10.1103/PhysRevE.80.046705}
}

@article{Wimsatt2025,
  author  = {Wimsatt, Gregory and Boyd, Alexander B. and Crutchfield, James P.},
  title   = {Trajectory Class Fluctuation Theorem},
  journal = {Journal of Statistical Physics},
  volume  = {192},
  pages   = {57},
  year    = {2025},
  doi     = {10.1007/s10955-025-03422-z},
  url     = {https://doi.org/10.1007/s10955-025-03422-z}
}

@article{KwonKim2019,
  author  = {Kwon, Hyukjoon and Kim, M. S.},
  title   = {Fluctuation Theorems for a Quantum Channel},
  journal = {Physical Review X},
  volume  = {9},
  pages   = {031029},
  year    = {2019},
  doi     = {10.1103/PhysRevX.9.031029},
  url     = {https://doi.org/10.1103/PhysRevX.9.031029}
}

@article{CuginiGuarnieri2026,
  author        = {Cugini, Davide and Guarnieri, Giacomo},
  title         = {State convertibility and fluctuation theorems from a dynamical reference: majorization meets martingales},
  year          = {2026},
  journal        = {arXiv: 2608.19391},
  url           = {https://arxiv.org/abs/2608.19391}
}

@article{Petz1994,
  author  = {Petz, D{\'e}nes},
  title   = {Geometry of canonical correlation on the state space of a quantum system},
  journal = {Journal of Mathematical Physics},
  volume  = {35},
  number  = {2},
  pages   = {780--795},
  year    = {1994},
  doi     = {10.1063/1.530611},
  url     = {https://doi.org/10.1063/1.530611}
}

@article{LesniewskiRuskai1999,
  author  = {Lesniewski, Andrew and Ruskai, Mary Beth},
  title   = {Monotone Riemannian metrics and relative entropy on noncommutative probability spaces},
  journal = {Journal of Mathematical Physics},
  volume  = {40},
  number  = {11},
  pages   = {5702--5724},
  year    = {1999},
  doi     = {10.1063/1.533053},
  url     = {https://doi.org/10.1063/1.533053}
}

@article{Scandi2019,
  author  = {Scandi, Matteo and Perarnau-Llobet, Mart{\'i}},
  title   = {Thermodynamic length in open quantum systems},
  journal = {Quantum},
  volume  = {3},
  pages   = {197},
  year    = {2019},
  doi     = {10.22331/q-2019-10-24-197},
  url     = {https://doi.org/10.22331/q-2019-10-24-197}
}

@article{Abiuso2020,
  author  = {Abiuso, Paolo and Miller, Harry J. D. and
             Perarnau-Llobet, Mart{\'i} and Scandi, Matteo},
  title   = {Geometric Optimisation of Quantum Thermodynamic Processes},
  journal = {Entropy},
  volume  = {22},
  number  = {10},
  pages   = {1076},
  year    = {2020},
  doi     = {10.3390/e22101076},
  url     = {https://doi.org/10.3390/e22101076}
}

@article{Minh2009,
  author  = {Minh, David D. L.},
  title   = {Density-Dependent Analysis of Nonequilibrium Paths Improves Free Energy Estimates},
  journal = {The Journal of Chemical Physics},
  volume  = {130},
  number  = {20},
  pages   = {204102},
  year    = {2009},
  doi     = {10.1063/1.3139189},
  url     = {https://doi.org/10.1063/1.3139189}
}

@article{MinhChodera2009,
  author  = {Minh, David D. L. and Chodera, John D.},
  title   = {Optimal Estimators and Asymptotic Variances for Nonequilibrium Path-Ensemble Averages},
  journal = {The Journal of Chemical Physics},
  volume  = {131},
  number  = {13},
  pages   = {134110},
  year    = {2009},
  doi     = {10.1063/1.3242285},
  url     = {https://doi.org/10.1063/1.3242285}
}

@article{MinhVaikuntanathan2011,
  author  = {Minh, David D. L. and Vaikuntanathan, Suriyanarayanan},
  title   = {Density-Dependent Analysis of Nonequilibrium Paths Improves Free Energy Estimates II. A Feynman--Kac Formalism},
  journal = {The Journal of Chemical Physics},
  volume  = {134},
  number  = {3},
  pages   = {034117},
  year    = {2011},
  doi     = {10.1063/1.3541152},
  url     = {https://doi.org/10.1063/1.3541152}
}

@article{AsbanRahav2017,
  author  = {Asban, Shahaf and Rahav, Saar},
  title   = {Nonequilibrium Free-Energy Estimation Conditioned on Measurement Outcomes},
  journal = {Physical Review E},
  volume  = {96},
  number  = {2},
  pages   = {022155},
  year    = {2017},
  doi     = {10.1103/PhysRevE.96.022155},
  url     = {https://doi.org/10.1103/PhysRevE.96.022155}
}

@article{NguyenMinh2020,
  author  = {Nguyen, Trung Hai and Minh, David D. L.},
  title   = {Implicit Ligand Theory for Relative Binding Free Energies: II. An Estimator Based on Control Variates},
  journal = {Journal of Physics Communications},
  volume  = {4},
  number  = {11},
  pages   = {115010},
  year    = {2020},
  doi     = {10.1088/2399-6528/abcbac},
  url     = {https://doi.org/10.1088/2399-6528/abcbac}
}

@inproceedings{DuEtAl2025FEAT,
  author    = {Du, Yuanqi and He, Jiajun and Vargas, Francisco and Wang, Yuanqing and Gomes, Carla P. and Hern{\'a}ndez-Lobato, Jos{\'e} Miguel and Vanden-Eijnden, Eric},
  title     = {{FEAT}: Free Energy Estimators with Adaptive Transport},
  booktitle = {Advances in Neural Information Processing Systems},
  volume    = {38},
  year      = {2025},
  doi       = {10.52202/085713-4756},
  url       = {https://papers.nips.cc/paper_files/paper/2025/hash/d0c3841867db2c516454845a450ca885-Abstract-Conference.html}
}

@article{LeeJarzynski2026,
  author  = {Lee, Sangyun and Jarzynski, Christopher},
  title   = {Estimating Free Energy Differences with Virtually Escorted Trajectories},
  journal = {arXiv:2606.30451},
  year    = {2026},
  url     = {https://arxiv.org/abs/2606.30451}
}

@article{BeyerGrubmuller2026,
  author  = {Beyer, David and Grubm{\"u}ller, Helmut},
  title   = {Optimal Intermediate Hamiltonians for Non-Equilibrium Free Energy Calculations: A Numerical Study of Markov Models},
  journal = {arXiv:2609.10519},
  year    = {2026},
  url     = {https://arxiv.org/abs/2609.10519}
}

@article{ScandiManzano2026,
  author  = {Scandi, Matteo and Manzano, Gonzalo},
  title   = {Universal Statistics of Charge Exchanges in Non-Abelian Quantum Transport},
  journal = {Physical Review Letters},
  volume  = {136},
  number  = {15},
  pages   = {150403},
  year    = {2026},
  doi     = {10.1103/n2gp-8bx9},
  url     = {https://doi.org/10.1103/n2gp-8bx9}
}

@article{RodriguesLutz2026,
  title = {Far-from-equilibrium thermodynamics of non-Abelian thermal states},
  author = {Rodrigues, Franklin L. S. and Lutz, Eric},
  journal = {Phys. Rev. Lett.},
  pages = {},
  year = {2026},
  month = {Aug},
  publisher = {American Physical Society},
  doi = {10.1103/jj8p-zmw2},
  url = {https://link.aps.org/doi/10.1103/jj8p-zmw2}
}

\clearpage
\onecolumngrid

\setcounter{section}{0}
\setcounter{subsection}{0}
\setcounter{equation}{0}
\setcounter{figure}{0}
\setcounter{table}{0}

\renewcommand{\thesection}{\arabic{section}}
\renewcommand{\thesubsection}{\Alph{subsection}}
\renewcommand{\theequation}{S\arabic{equation}}
\renewcommand{\thefigure}{S\arabic{figure}}
\renewcommand{\thetable}{S\arabic{table}}

\makeatletter
\renewcommand{\p@subsection}{\thesection}
\makeatother

\newcommand{\suppnote}[1]{%
    \refstepcounter{section}%
    \setcounter{subsection}{0}%
    \section*{Supplementary Note \arabic{section}. #1}%
    \addcontentsline{toc}{section}{Supplementary Note \arabic{section}. #1}%
}

\newcommand{\suppsubsection}[1]{%
    \refstepcounter{subsection}%
    \subsection*{\Alph{subsection}. #1}%
    \addcontentsline{toc}{subsection}{\Alph{subsection}. #1}%
}

\section*{Supplemental Material for\\
Physical-Work Fluctuation Relations from Accessible Quantum Macrostates}

This Supplemental Material follows the order of the main text and keeps each technical step separate.
Note~\ref{supp:note-targetB-derivation} derives the physical-work-and-record fluctuation relation, explains the two noncommuting-measurement corrections, and connects the unresolved endpoint information to the finite-step return construction of Ref.~\cite{Ahmadi2026Endogenous}.
Note~\ref{supp:note-BH-validation} gives the Bose--Hubbard model, the time-resolved endpoint fits, and independent numerical checks.
Note~\ref{supp:note-sampling-frontier} turns the exact family into control variables for one common Jarzynski free-energy target, defines the finite-confidence benchmark, and derives the information--sampling frontier.
Note~\ref{supp:note-arbitrary-initial} extends the fluctuation relation beyond canonical preparation and explains the extra closure, support, and calibration requirements.

\suppnote{Exact physical-work-and-record fluctuation relation}
\label{supp:note-targetB-derivation}

\suppsubsection{Setup and measurement sequence}

We work in a finite-dimensional Hilbert space.
The initial and final Hamiltonians have spectral decompositions
\begin{equation}
H_i
=
\sum_n
E_n^i\Pi_n^i,
\quad
H_f
=
\sum_m
E_m^f\Pi_m^f.
\label{eq:supp-Hi-Hf}
\end{equation}
The projectors may have arbitrary degeneracies.
The initial state is canonical,
\begin{equation}
\rho_i
=
\frac{
e^{-\beta_iH_i}
}{
Z_i
},
\quad
Z_i
=
\Tr
e^{-\beta_iH_i}.
\label{eq:supp-rhoi}
\end{equation}
An isolated control protocol produces a unitary \(U\).
Before any final measurement is made, the microscopic endpoint is
\begin{equation}
\rho_f
=
U\rho_iU^\dagger.
\label{eq:supp-rhof}
\end{equation}

The thermodynamic observer does not retain the whole matrix \(\rho_f\).
Instead, the retained endpoint data are the mean energy
\(E_f=\Tr(H_f\rho_f)\)
and the probabilities
\(p_{r,f}=\Tr(P_r\rho_f)\)
of a projective record,
\begin{equation}
P_rP_s
=
\delta_{rs}P_r,
\quad
\sum_rP_r
=
I.
\label{eq:supp-record-pvm}
\end{equation}
The maximum-entropy (MaxEnt) state compatible with these data is
\begin{equation}
\bar\rho_f
=
\frac{
e^{-\beta_f^\star H_f-A_f}
}{
Z_f
},
\quad
A_f
=
\sum_r
\alpha_{r,f}P_r,
\quad
Z_f
=
\Tr
e^{-\beta_f^\star H_f-A_f}.
\label{eq:supp-rhobar}
\end{equation}
The multiplier \(\beta_f^\star\) is fixed by the mean-energy constraint.
Away from equilibrium it is a maximum-entropy multiplier and is not assumed to be a physical inverse temperature.

The endpoint parameters in Eq.~\eqref{eq:supp-rhobar} are calibrated on copies that are not subjected to the fluctuation measurement sequence.
After this calibration, the parameters are frozen.
A fluctuation run then resolves
\begin{equation}
H_i
\longrightarrow
U
\longrightarrow
H_f
\longrightarrow
R_f.
\label{eq:supp-sequence}
\end{equation}
The first two measurements define the usual two-point-measurement work,
\begin{equation}
W_{mn}
=
E_m^f-E_n^i.
\label{eq:supp-work}
\end{equation}

The role of TPM is deliberately narrow.
For the unmeasured isolated process, the mean work transferred through the external control is already well defined,
\begin{equation}
\overline W_{\rm on}
=
\int_0^\tau
dt\,
\Tr
\left[
\rho_t\dot H_t
\right]
=
\Tr
\left[
H_f\rho_f
\right]
-
\Tr
\left[
H_i\rho_i
\right].
\label{eq:supp-mean-work}
\end{equation}
TPM is introduced only because the fluctuation relation needs a realization-level random variable.
For the canonical state in Eq.~\eqref{eq:supp-rhoi}, the first nonselective energy measurement leaves the ensemble state unchanged because
\([\rho_i,H_i]=0\).

The joint probability of obtaining the initial energy \(E_n^i\), final energy \(E_m^f\), and final record value \(r\) is
\begin{equation}
p(n,m,r)
=
\frac{
e^{-\beta_iE_n^i}
}{
Z_i
}
\Tr
\left[
P_r
\Pi_m^f
U
\Pi_n^i
U^\dagger
\Pi_m^f
P_r
\right].
\label{eq:supp-joint}
\end{equation}
No commutation relation between \(H_f\) and \(P_r\) is used here.
For a degenerate initial energy, the factor
\(e^{-\beta_iE_n^i}/Z_i\)
is the eigenvalue of the canonical density operator inside the whole eigenspace \(\Pi_n^i\).
The trace in Eq.~\eqref{eq:supp-joint} automatically accounts for its dimension.

Using the completeness of both measurements and the unitarity of \(U\) gives
\begin{equation}
\sum_{n,m,r}
p(n,m,r)
=
1.
\label{eq:supp-joint-normalized}
\end{equation}
Thus Eq.~\eqref{eq:supp-joint} is a genuine positive trajectory distribution.
Its trace factor has a direct meaning: the unitary carries the selected initial energy sector to the final energy sector, after which the record measurement resolves how much of that sector lies in outcome \(r\).

\suppsubsection{Exact exponential average}

We first define
\begin{equation}
\zeta_{nmr}
=
\beta_iW_{mn}
+
\left(
\beta_f^\star-\beta_i
\right)
E_m^f
+
\alpha_{r,f}.
\label{eq:supp-zeta}
\end{equation}
Substituting Eq.~\eqref{eq:supp-work} into the exponential weight gives
\begin{equation}
e^{-\zeta_{nmr}}
=
e^{\beta_iE_n^i}
e^{-\beta_f^\star E_m^f}
e^{-\alpha_{r,f}}.
\label{eq:supp-zeta-exp}
\end{equation}
The factor \(e^{\beta_iE_n^i}\) cancels the canonical weight in Eq.~\eqref{eq:supp-joint}.
This cancellation is the same basic step that makes the ordinary Jarzynski relation possible.
The exponential average therefore becomes
\begin{equation}
\begin{aligned}
\left\langle
e^{-\zeta}
\right\rangle
=
\frac{1}{Z_i}
\sum_{n,m,r}
e^{-\beta_f^\star E_m^f}
e^{-\alpha_{r,f}}
\Tr
\left[
P_r
\Pi_m^f
U
\Pi_n^i
U^\dagger
\Pi_m^f
P_r
\right].
\end{aligned}
\label{eq:supp-zeta-average-1}
\end{equation}
Summing over the initial energy projectors gives
\begin{equation}
\sum_n
U\Pi_n^iU^\dagger
=
I.
\label{eq:supp-unitarity-sum}
\end{equation}
The detailed unitary then disappears from the average,
\begin{equation}
\left\langle
e^{-\zeta}
\right\rangle
=
\frac{1}{Z_i}
\sum_{m,r}
e^{-\beta_f^\star E_m^f}
e^{-\alpha_{r,f}}
\Tr
\left[
\Pi_m^fP_r
\right].
\label{eq:supp-zeta-average-2}
\end{equation}

Since
\begin{equation}
e^{-\beta_f^\star H_f}
=
\sum_m
e^{-\beta_f^\star E_m^f}
\Pi_m^f,
\quad
e^{-A_f}
=
\sum_r
e^{-\alpha_{r,f}}P_r,
\label{eq:supp-expansions}
\end{equation}
Eq.~\eqref{eq:supp-zeta-average-2} can be written as
\begin{equation}
\left\langle
e^{-\zeta}
\right\rangle
=
\frac{
Z_f^{\rm seq}
}{
Z_i
},
\label{eq:supp-zeta-identity}
\end{equation}
where
\begin{equation}
Z_f^{\rm seq}
=
\Tr
\left[
e^{-\beta_f^\star H_f}
e^{-A_f}
\right].
\label{eq:supp-Zseq}
\end{equation}
When \(H_f\) and \(A_f\) do not commute, the product in Eq.~\eqref{eq:supp-Zseq} is generally different from the joint exponential \(e^{-(\beta_f^\star H_f+A_f)}\).
Because the trace is cyclic, reversing the two displayed factors alone does not change the scalar \(Z_f^{\rm seq}\); the physical measurement order is shown more directly by the dephasing form below.

The final energy measurement can also be displayed explicitly.
Define the energy-dephasing map
\begin{equation}
\cD_{H_f}[X]
=
\sum_m
\Pi_m^f
X
\Pi_m^f.
\label{eq:supp-final-dephasing}
\end{equation}
Because \(e^{-\beta_f^\star H_f}\) is invariant under this map,
\begin{equation}
Z_f^{\rm seq}
=
\Tr
\left[
e^{-\beta_f^\star H_f}
\cD_{H_f}
\left[
e^{-A_f}
\right]
\right].
\label{eq:supp-Zseq-dephased}
\end{equation}
Thus Eq.~\eqref{eq:supp-Zseq-dephased} makes explicit the physical dephasing caused by measuring the final energy before the record.

\suppsubsection{Endpoint noncommutativity and the exact IFT}

The thermodynamic normalization \(Z_f\) and the sequential normalization \(Z_f^{\rm seq}\) need not agree.
We define
\begin{equation}
\mathcal Q_{\rm nc}
=
\ln
\frac{
Z_f^{\rm seq}
}{
Z_f
}.
\label{eq:supp-Qnc}
\end{equation}
The Golden--Thompson inequality gives
\begin{equation}
\Tr
e^{-(\beta_f^\star H_f+A_f)}
\leq
\Tr
\left[
e^{-\beta_f^\star H_f}
e^{-A_f}
\right],
\label{eq:supp-GT}
\end{equation}
and therefore
\begin{equation}
\mathcal Q_{\rm nc}
\geq
0.
\label{eq:supp-Qnc-positive}
\end{equation}
For finite-dimensional Hermitian operators, equality holds when
\([\beta_f^\star H_f,A_f]=0\).
For \(\beta_f^\star\neq0\), this is equivalent to
\([H_f,A_f]=0\).

For compactness, define the dimensionless normalization potentials
\begin{equation}
\Psi_i
=
-\ln Z_i,
\quad
\Psi_f
=
-\ln Z_f,
\quad
\Delta\Psi
=
\Psi_f-\Psi_i.
\label{eq:supp-Psi}
\end{equation}
For the canonical initial state \(\Psi_i=\beta_iF_i\), whereas \(\Psi_f\) is only the logarithmic normalization of the nonequilibrium MaxEnt representative and is not assumed to be an equilibrium free energy.
Using
\(Z_f^{\rm seq}=Z_fe^{\mathcal Q_{\rm nc}}\),
Eq.~\eqref{eq:supp-zeta-identity} becomes
\begin{equation}
\left\langle
\exp
\left[
-\beta_iW
-
\left(
\beta_f^\star-\beta_i
\right)
E_m^f
-
\alpha_{r,f}
\right]
\right\rangle
=
e^{-\Delta\Psi+\mathcal Q_{\rm nc}}.
\label{eq:supp-main-unshifted}
\end{equation}
It is convenient to absorb the endpoint normalization into the trajectory variable,
\begin{equation}
\sigma_{\rm phys,R}
=
\beta_iW
+
\left(
\beta_f^\star-\beta_i
\right)
E_m^f
+
\alpha_{r,f}
+
\mathcal Q_{\rm nc}
-
\Delta\Psi.
\label{eq:supp-sigma}
\end{equation}
The exact integral fluctuation relation is then
\begin{equation}
\left\langle
e^{-\sigma_{\rm phys,R}}
\right\rangle
=
1.
\label{eq:supp-IFT}
\end{equation}
We refer to Eq.~\eqref{eq:supp-IFT} as the integral fluctuation theorem (IFT).
The quantity \(\sigma_{\rm phys,R}\) is not work.
Its first term contains the ordinary TPM work, while the remaining terms supply the information required by the chosen endpoint macrostate.

A common shift of all record multipliers,
\(\alpha_{r,f}\rightarrow\alpha_{r,f}+c\),
changes both \(Z_f\) and \(Z_f^{\rm seq}\) by the same factor.
As a result, \(\mathcal Q_{\rm nc}\) is unchanged and the shift of \(\alpha_{r,f}\) is canceled by the shift of \(\Psi_f\).
Thus \(\sigma_{\rm phys,R}\) is independent of the arbitrary additive gauge of the record multipliers.

If the record is removed by setting \(\alpha_{r,f}=0\) and the final multiplier is chosen as \(\beta_f^\star=\beta_i\), then
\(\mathcal Q_{\rm nc}=0\) and Eq.~\eqref{eq:supp-IFT} reduces to the usual Jarzynski form,
\begin{equation}
\left\langle
e^{-\beta_i(W-\Delta F)}
\right\rangle
=
1.
\label{eq:supp-Jarzynski-limit}
\end{equation}
The present relation therefore keeps the standard work convention and adds the information carried by the retained quantum macrostate.
The comparison also makes the operational difference precise: ordinary Jarzynski uses microscopic work trajectories to recover the canonical equilibrium reference fixed by \(\beta_i\) and \(H_f\), whereas the thermodynamic member above keeps the same fine-grained work labels but uses the MaxEnt reference fitted to the actual undisturbed endpoint energy and record.

\suppsubsection{Measurement ordering and the mean relation}

The thermodynamic endpoint is defined from the undisturbed state \(\rho_f\).
Its record probabilities are
\begin{equation}
p_{r,f}
=
\Tr
\left[
P_r\rho_f
\right].
\label{eq:supp-p-undisturbed}
\end{equation}
In a fluctuation run, however, the record is measured after the final energy.
Its distribution is therefore
\begin{equation}
p_{r,f}^{\rm seq}
=
\Tr
\left[
P_r
\cD_{H_f}[\rho_f]
\right].
\label{eq:supp-p-seq}
\end{equation}
The two distributions coincide when the relevant record is unaffected by the energy dephasing, but they are different in general.
Averaging Eq.~\eqref{eq:supp-sigma} gives
\begin{equation}
\begin{aligned}
\left\langle
\sigma_{\rm phys,R}
\right\rangle
=
\beta_i
\left\langle W\right\rangle
+
\left(
\beta_f^\star-\beta_i
\right)
E_f
+
\sum_r
\alpha_{r,f}
p_{r,f}^{\rm seq}
+
\mathcal Q_{\rm nc}
-
\Delta\Psi.
\end{aligned}
\label{eq:supp-mean-direct}
\end{equation}
For the canonical initial state, unitary evolution preserves the von Neumann entropy and
\(\langle W\rangle=E_f-E_i\).
Using the endpoint MaxEnt state then gives
\begin{equation}
D
\left(
\rho_f
\Vert
\bar\rho_f
\right)
=
\beta_i
\left\langle W\right\rangle
+
\left(
\beta_f^\star-\beta_i
\right)
E_f
+
\sum_r
\alpha_{r,f}
p_{r,f}
-
\Delta\Psi.
\label{eq:supp-D-expression}
\end{equation}
Subtracting Eq.~\eqref{eq:supp-D-expression} from Eq.~\eqref{eq:supp-mean-direct} gives
\begin{equation}
\left\langle
\sigma_{\rm phys,R}
\right\rangle
=
D
\left(
\rho_f
\Vert
\bar\rho_f
\right)
+
\Delta_{\rm meas}
+
\mathcal Q_{\rm nc},
\label{eq:supp-mean-decomposition}
\end{equation}
with
\begin{equation}
\Delta_{\rm meas}
=
\sum_r
\alpha_{r,f}
\left(
p_{r,f}^{\rm seq}
-
p_{r,f}
\right).
\label{eq:supp-Delta-meas}
\end{equation}
Equivalently,
\begin{equation}
\Delta_{\rm meas}
=
\Tr
\left[
A_f
\left(
\cD_{H_f}[\rho_f]-\rho_f
\right)
\right].
\label{eq:supp-Delta-operator}
\end{equation}
The two correction terms have different meanings.
The scalar \(\mathcal Q_{\rm nc}\) compares two endpoint normalizations and is nonnegative.
The quantity \(\Delta_{\rm meas}\) measures how the energy measurement changes the weighted record and has no fixed sign.
Because energy dephasing leaves every operator that commutes with \(H_f\) unchanged, \(\Delta_{\rm meas}\) vanishes together with \(\mathcal Q_{\rm nc}\) when \([H_f,A_f]=0\).
The complete mean remains nonnegative by Jensen's inequality applied to Eq.~\eqref{eq:supp-IFT}.

For the canonical isolated work stroke used in the main text, denoted \(B\rightarrow C^{-}\),
\begin{equation}
D
\left(
\rho_{C^-}
\Vert
\bar\rho_{E,R}(C^-)
\right)
=
\frac{
\Sigma_{BC}
}{
\kB
}.
\label{eq:supp-D-Sigma}
\end{equation}
Equation~\eqref{eq:supp-mean-decomposition} therefore tells us exactly which part of the mean fluctuation variable comes from the thermodynamic information loss and which part is introduced by the sequential measurement itself.

\suppsubsection{Unresolved microscopic structure and dynamical return}

The first term in Eq.~\eqref{eq:supp-mean-decomposition} has a direct connection to the unresolved-state construction of Ref.~\cite{Ahmadi2026Endogenous}.
The record-only theory decomposes the exact state as \(\rho_t=\mathcal G[\rho_t]+\chi_t\), where \(\mathcal G[\rho_t]\) reproduces the present record and \(\chi_t\) is invisible to it.
After a finite unitary step \(U_{t+\tau,t}\), the representative alone predicts \(K^{(U)}(t,\tau)p(t)\), while the hidden part supplies the correction. The exact future record is therefore
\begin{equation}
p(t+\tau)
=
K^{(U)}(t,\tau)p(t)
+
\mathbf r(t,\tau),
\label{eq:supp-foundational-return}
\end{equation}
with components
\begin{equation}
r_s(t,\tau)
=
\Tr\!\left[
P_sU_{t+\tau,t}\chi_tU_{t+\tau,t}^\dagger
\right].
\label{eq:supp-foundational-return-component}
\end{equation}
Thus \(\chi_t\) is the hidden microscopic structure at the present time, whereas \(\mathbf r(t,\tau)\) is the part of a later accessible record generated by that structure.

The present thermodynamic state retains more information than the record-only representative because it fixes both the mean energy and the record probabilities.
The corresponding unresolved operator is therefore
\begin{equation}
\chi_{E,R}(t)
=
\rho_t
-
\bar\rho_{E,R}(t).
\label{eq:supp-chi-ER}
\end{equation}
Since \(\bar\rho_{E,R}(t)\) reproduces all retained constraints,
\begin{equation}
\Tr\chi_{E,R}(t)
=
0,
\quad
\Tr\!\left[H_t\chi_{E,R}(t)\right]
=
0,
\quad
\Tr\!\left[P_s\chi_{E,R}(t)\right]
=
0
\quad
\forall s.
\label{eq:supp-chi-ER-invisible}
\end{equation}
The operator \(\chi_{E,R}(t)\) is therefore invisible to the complete thermodynamic state used in this work at the instant \(t\).
This does not make it dynamically irrelevant.
For any later endpoint, the exact spatial record can be written as
\begin{equation}
p_s(t+\tau)
=
\bar p_s(t,\tau)
+
r_s^{(E,R)}(t,\tau),
\label{eq:supp-ER-record-return}
\end{equation}
where
\begin{equation}
\bar p_s(t,\tau)
=
\Tr\!\left[
P_sU_{t+\tau,t}\bar\rho_{E,R}(t)U_{t+\tau,t}^\dagger
\right],
\label{eq:supp-ER-record-baseline}
\end{equation}
and
\begin{equation}
r_s^{(E,R)}(t,\tau)
=
\Tr\!\left[
P_sU_{t+\tau,t}\chi_{E,R}(t)U_{t+\tau,t}^\dagger
\right].
\label{eq:supp-ER-record-return-component}
\end{equation}
The same split holds for the retained energy,
\begin{equation}
E(t+\tau)
=
\bar E(t,\tau)
+
r_E^{(E,R)}(t,\tau),
\label{eq:supp-ER-energy-return}
\end{equation}
with
\begin{equation}
\begin{aligned}
\bar E(t,\tau)
&=
\Tr\!\left[
H_{t+\tau}U_{t+\tau,t}\bar\rho_{E,R}(t)U_{t+\tau,t}^\dagger
\right],
\\
r_E^{(E,R)}(t,\tau)
&=
\Tr\!\left[
H_{t+\tau}U_{t+\tau,t}\chi_{E,R}(t)U_{t+\tau,t}^\dagger
\right].
\end{aligned}
\label{eq:supp-ER-energy-return-components}
\end{equation}
At \(\tau=0\), all return contributions in Eqs.~\eqref{eq:supp-ER-record-return-component} and \eqref{eq:supp-ER-energy-return-components} vanish because the representative matches the retained energy and record at time \(t\).
At later times they need not vanish.
When the representative is specialized back to the record-only map \(\mathcal G\), Eq.~\eqref{eq:supp-ER-record-return} reduces to the finite-step structure in Eq.~\eqref{eq:supp-foundational-return}, with the representative contribution becoming \(K^{(U)}p(t)\).

The relative entropy used in the present fluctuation theorem and the return term therefore describe complementary aspects of the same unresolved sector.
The quantity \(D(\rho_t\Vert\bar\rho_{E,R}(t))\) is a state-level measure of how much microscopic information is missing from the thermodynamic representative; it vanishes exactly when \(\chi_{E,R}(t)=0\).
The return vector \(\mathbf r^{(E,R)}(t,\tau)\), by contrast, is dynamical and protocol dependent: it records how much of that hidden structure becomes visible in a specified future record.
A nonzero information gap does not by itself determine the magnitude or sign of a later return.

This distinction also clarifies Eq.~\eqref{eq:supp-mean-decomposition}.
Its relative-entropy term is the information-theoretic contribution associated with the unresolved state \(\chi_{E,R}(f)\).
The terms \(\Delta_{\rm meas}\) and \(\mathcal Q_{\rm nc}\) are not return terms: they arise because the fluctuation protocol measures the final energy before the independent thermodynamic record when these two observables need not commute.
The fluctuation relation therefore separates the pre-existing microscopic information omitted by the thermodynamic state from additional effects created by the measurement sequence.

For the canonical Bose--Hubbard preparation used in the main text, \(\rho_0=\bar\rho_{E,R}(0)\), so \(\chi_{E,R}(0)=0\) and the unresolved relative entropy starts from zero.
During the isolated drive, \(\rho_t\) develops structure that is not fixed by the instantaneous energy and spatial record, and \(D(\rho_t\Vert\bar\rho_{E,R}(t))\) grows accordingly.
The arbitrary-state construction below gives an explicit complementary example in which two preparations share the same initial \((E,R)\) state but have different unresolved components; the difference then reappears in their future records exactly through the return structure above.

\suppnote{Bose--Hubbard implementation and numerical validation}
\label{supp:note-BH-validation}

\suppsubsection{Working medium and retained record}

We use four bosons on a four-site open Bose--Hubbard chain.
Throughout the numerics,
\(J=\hbar=\kB=1\).
The Hamiltonian is
\begin{equation}
H(\Lambda)
=
-J
\sum_{j=1}^{L-1}
\left(
a_j^\dagger a_{j+1}
+
a_{j+1}^\dagger a_j
\right)
+
\frac{U}{2}
\sum_{j=1}^{L}
n_j(n_j-1)
+
\Lambda N_R,
\label{eq:supp-BH}
\end{equation}
Here \(n_j=a_j^\dagger a_j\), and
\begin{equation}
N_R
=
n_3+n_4.
\label{eq:supp-NR}
\end{equation}
The hopping term moves bosons between neighboring sites, the interaction term assigns an energy cost to multiple occupation of one site, and the last term raises or lowers the right half relative to the left.
The total particle number is fixed at \(Q=4\).
With \(L=4\), the fixed-particle Hilbert-space dimension is
\begin{equation}
d
=
\binom{Q+L-1}{Q}
=
35.
\label{eq:supp-dim}
\end{equation}

The retained spatial record is the number of particles on the right half,
\begin{equation}
r
=
n_R
\in
\{0,1,2,3,4\}.
\label{eq:supp-r-values}
\end{equation}
In the Fock basis,
\begin{equation}
P_r
=
\sum_{\substack{
n_1+n_2+n_3+n_4=Q\\
n_3+n_4=r
}}
\left|
n_1,n_2,n_3,n_4
\right\rangle
\left\langle
n_1,n_2,n_3,n_4
\right|.
\label{eq:supp-Pr}
\end{equation}
The thermodynamic record is therefore the mean energy together with the five probabilities \(p_r\), of which four are independent after normalization.
The initial state is canonical at
\(\beta_iJ=0.2\)
with
\(\Lambda_0/J=20\).
The interaction and protocol duration are
\begin{equation}
\frac{U}{J}
=
\frac{1.3}{3},
\quad
J\tau
=
20.
\label{eq:supp-parameters}
\end{equation}
During the isolated work stroke,
\begin{equation}
\Lambda(t)
=
\Lambda_0
\left(
1-\frac{t}{\tau}
\right),
\quad
0
\leq
t
\leq
\tau.
\label{eq:supp-ramp}
\end{equation}
Thus the large initial energy offset is removed continuously while the system remains isolated.

This is the \(B\rightarrow C^-\) work stroke referred to in the main text and inherited from Ref.~\cite{Ahmadi2026FiniteParticle}.
Its definition is local and does not require the rest of that thermodynamic cycle.
The starting point is
\begin{equation}
H_B
=
H(\Lambda_0),
\quad
\rho_B
=
\frac{e^{-\beta_i H_B}}{Z_B},
\end{equation}
and the final physical Hamiltonian is
\(H_{C^-}=H(0)\).
During the ramp the working-medium boundary is thermally isolated, so \(\delta Q=0\); energy is transferred only as work through the controlled parameter \(\Lambda\).
Since \(\partial H/\partial\Lambda=N_R\), the mean work done on the working medium is
\begin{equation}
\overline W_{\rm on}^{B\to C^-}
=
\int_0^\tau
 dt\,
\dot\Lambda(t)
\Tr\!\left[N_R\rho_t\right]
=
E_{C^-}-E_B.
\label{eq:supp-BC-work}
\end{equation}
Because \(\Lambda\) is lowered, the nominal protocol gives negative work on the working medium, i.e., work extraction.
The exact endpoint before any final measurement is
\begin{equation}
\rho_{C^-}
=
U_{BC}\rho_BU_{BC}^\dagger,
\end{equation}
and is generally not a Gibbs state of \(H(0)\).
The notation \(C^-\) means immediately after this isolated ramp and before the subsequent work-only Hamiltonian switch and cold-side continuation used to complete the cycle in Ref.~\cite{Ahmadi2026FiniteParticle}.
Those later operations play no role here: the present theorem and all reported trajectory statistics concern only the isolated \(B\rightarrow C^-\) process.
In the generic notation of Note~\ref{supp:note-targetB-derivation}, \(H_i=H_B\), \(H_f=H_{C^-}\), \(\rho_i=\rho_B\), and \(\rho_f=\rho_{C^-}\).

At every chosen endpoint \(t\), we compute the exact microscopic state
\(\rho_t=U_t\rho_iU_t^\dagger\),
its mean energy, and its record probabilities.
We then fit
\begin{equation}
\bar\rho_t
=
\frac{
\exp
\left[
-\beta^\star_tH_t
-
\sum_r
\alpha_r(t)P_r
\right]
}{
Z_t
}
\label{eq:supp-time-maxent}
\end{equation}
to the same energy and record.
We fix the additive gauge by setting
\begin{equation}
\alpha_4(t)
=
0.
\label{eq:supp-gauge}
\end{equation}
The remaining five independent parameters are determined from the mean energy and \(p_0,\ldots,p_3\).

Because \(\rho_t\) and \(\bar\rho_t\) have the same retained constraints,
their relative entropy can also be written as the entropy gained when the unresolved microscopic information is discarded,
\begin{equation}
D
\left(
\rho_t
\Vert
\bar\rho_t
\right)
=
S(\bar\rho_t)
-
S(\rho_t).
\label{eq:supp-D-entropy}
\end{equation}
The microscopic entropy is constant under the isolated unitary.
The growth of Eq.~\eqref{eq:supp-D-entropy} therefore comes entirely from information that becomes invisible to the retained thermodynamic state.

\begin{table}[t]
\caption{Nominal Bose--Hubbard parameters and endpoint quantities.The gauge is fixed by \(\alpha_4=0\).}
\label{tab:supp-model-endpoint}

\centering
\begin{minipage}{0.83\columnwidth}
\centering

\begin{ruledtabular}
\begin{tabular}{lc}
Quantity & Value \\
\hline
Hilbert-space dimension & \(35\) \\
\(L,Q\) & \(4,4\) \\
\(U/J\) & \(0.433333333\) \\
\(\Lambda_0/J\) & \(20\) \\
\(\beta_iJ\) & \(0.2\) \\
\(J\tau\) & \(20\) \\
\(E_B/J\) & \(0.571302382040\) \\
\(S_B/\kB\) & \(1.603917951434\) \\
\(E_{C^-}/J\) & \(-3.801854896100\) \\
\(W_{\rm on}/J\) & \(-4.373157278140\) \\
\(\beta_f^\star J\) & \(0.626142210199\) \\
\(D(\rho_f\Vert\bar\rho_f)\) & \(0.641609555870\) \\
\(\Delta_{\rm meas}\) & \(0.031325474797\) \\
\(\mathcal Q_{\rm nc}\) & \(0.001878908345\) \\
\(\langle\sigma_{\rm phys,R}\rangle\) & \(0.674813939012\) \\
\end{tabular}
\end{ruledtabular}

\end{minipage}
\end{table}

Table~\ref{tab:supp-model-endpoint} collects the nominal parameters and the endpoint quantities used throughout the numerical checks.
At the final point, the undisturbed record is
\begin{equation}
\mathbf p_f
=
\left(
0.1290675,
0.2912160,
0.2993206,
0.1901327,
0.0902632
\right),
\label{eq:supp-p-final}
\end{equation}
while the sequential record is
\begin{equation}
\mathbf p_f^{\rm seq}
=
\left(
0.1100533,
0.2458246,
0.2882442,
0.2458246,
0.1100533
\right).
\label{eq:supp-pseq-final}
\end{equation}
The fitted record multipliers are
\begin{equation}
\boldsymbol{\alpha}_f
=
\left(
-0.3439218,
-0.3024354,
-0.0516551,
0.1882853,
0
\right).
\label{eq:supp-alpha-final}
\end{equation}

The symmetry of Eq.~\eqref{eq:supp-pseq-final} has a simple origin.
At \(\Lambda=0\), the Hamiltonian is invariant under left--right reflection.
Its spectrum is nondegenerate in the working sector, so its energy eigenstates can be chosen with definite reflection parity.
After energy dephasing, the state is a mixture of these parity eigenstates and therefore gives equal probabilities for \(r\) and \(Q-r\).
The undisturbed endpoint still contains energy-basis coherences and need not have this symmetry.

\suppsubsection{Time-resolved behavior}

\begin{figure}[t]
\includegraphics[width=0.8\textwidth]{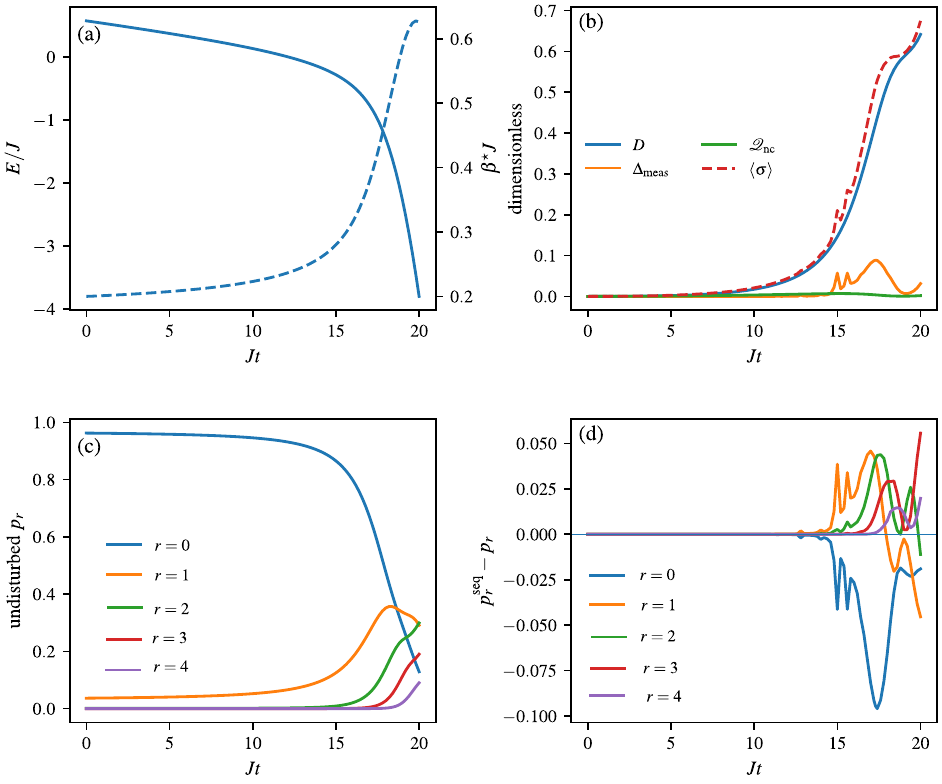}
\caption{
Time-resolved nominal Bose--Hubbard protocol.
(a) Mean energy \(E/J\) and fitted maximum-entropy multiplier \(\beta^\star J\).
The energy falls as the right-half offset is removed, while the multiplier adjusts so that the representative keeps the measured energy and record at each endpoint.
(b) Unresolved information \(D\), measurement correction \(\Delta_{\rm meas}\), noncommutativity gap \(\mathcal Q_{\rm nc}\), and their combination \(\langle\sigma_{\rm phys,R}\rangle=D+\Delta_{\rm meas}+\mathcal Q_{\rm nc}\).
(c) Undisturbed spatial-record probabilities \(p_r\).
(d) Change of each record probability caused by the final-energy dephasing, \(p_r^{\rm seq}-p_r\).
Each time is evaluated on a separate ensemble of copies.
}
\label{fig:supp-time}
\end{figure}

Figure~\ref{fig:supp-time} shows the complete time-resolved nominal protocol.
Each value of \(t\) is a separate truncated protocol performed on independently prepared copies.
The curves should therefore not be read as a sequence of invasive measurements on one realization.

Figure~\figpanel{fig:supp-time}{a} shows that the mean energy decreases throughout the work stroke while the fitted \(\beta^\star\) changes continuously.
The multiplier follows the nonequilibrium endpoint constraints and should not be interpreted as a physical temperature.

Figure~\figpanel{fig:supp-time}{b} separates the thermodynamic and measurement contributions.
The unresolved information \(D\) starts from zero because the initial canonical state is already the maximum-entropy state compatible with its own energy and record, so \(\chi_{E,R}(0)=0\).
It then grows as \(\chi_{E,R}(t)\) develops microscopic structure that is not fixed by the retained data.
The mean fluctuation variable follows the same overall growth but also contains the two measurement corrections.
The noncommutativity gap reaches its largest value,
\begin{equation}
\mathcal Q_{\rm nc}^{\rm max}
=
0.00693753,
\quad
Jt
=
15.2,
\label{eq:supp-Qnc-max}
\end{equation}
while the measurement-order correction reaches
\begin{equation}
\Delta_{\rm meas}^{\rm max}
=
0.0886154,
\quad
Jt
=
17.3.
\label{eq:supp-Delta-max}
\end{equation}
Their maxima occur at different times because they measure different effects.

Figure~\figpanel{fig:supp-time}{c} shows the redistribution of particles between the two halves of the chain.
Figure~\figpanel{fig:supp-time}{d} shows when the final energy measurement begins to change that spatial information appreciably.
The late-time growth of these differences explains why \(\Delta_{\rm meas}\) becomes important near the end of the stroke.

\suppsubsection{Independent propagation and MaxEnt checks}

The main propagation uses an adaptive DOP853 integrator with relative and absolute tolerances \(10^{-11}\) and \(10^{-13}\).
We independently repeat the endpoint propagation with a fourth-order two-node Gauss--Magnus scheme using \(J\Delta t=0.01\).

For one time step, define
\begin{equation}
t_{1,2}
=
t
+
\left(
\frac{1}{2}
\mp
\frac{\sqrt{3}}{6}
\right)
\Delta t,
\quad
H_{1,2}
=
H(t_{1,2}).
\label{eq:supp-magnus-nodes}
\end{equation}
The fourth-order Magnus exponent is
\begin{equation}
\Omega
=
-\frac{i\Delta t}{2}
\left(
H_1+H_2
\right)
+
\frac{
\sqrt{3}\Delta t^2
}{
12
}
\left[
H_1,H_2
\right],
\label{eq:supp-magnus}
\end{equation}
and the step propagator is \(e^\Omega\).
The final DOP853 and Magnus propagators differ in operator norm by
\begin{equation}
\left\|
U_{\rm DOP853}
-
U_{\rm Magnus}
\right\|_\infty
=
4.37\times10^{-9}.
\label{eq:supp-U-difference}
\end{equation}

\begin{table}[t]
\caption{
Independent numerical validation of the nominal endpoint.
The two propagation methods agree far beyond the precision needed for the figures.
}
\label{tab:supp-numerical-validation}
\begin{ruledtabular}
\begin{tabular}{lccc}
Quantity & DOP853 & Magnus & Absolute difference \\
\hline
\(E_f/J\)
& \(-3.801854896100\)
& \(-3.801854896180\)
& \(8.01\times10^{-11}\) \\
\(D\)
& \(0.641609555870\)
& \(0.641609555817\)
& \(5.33\times10^{-11}\) \\
\(\Delta_{\rm meas}\)
& \(0.0313254747970\)
& \(0.0313254748035\)
& \(6.46\times10^{-12}\) \\
\(\mathcal Q_{\rm nc}\)
& \(0.00187890834451\)
& \(0.00187890834465\)
& \(1.32\times10^{-13}\) \\
\(\langle\sigma\rangle\)
& \(0.674813939012\)
& \(0.674813938965\)
& \(4.67\times10^{-11}\) \\
\(\langle e^{-\sigma}\rangle\)
& \(1.000000000000005\)
& \(1.000000000000003\)
& \(2.22\times10^{-15}\) \\
\end{tabular}
\end{ruledtabular}
\end{table}

The largest MaxEnt constraint residual over the full time scan is
\begin{equation}
5.78\times10^{-10}.
\label{eq:supp-maxent-residual}
\end{equation}
The agreement in Table~\ref{tab:supp-numerical-validation} shows that the endpoint decomposition and the exact IFT are not artifacts of one time-propagation method.

\suppnote{Exact family as thermodynamic controls for free-energy estimation}
\label{supp:note-sampling-frontier}

\suppsubsection{Exact family and the common Jarzynski target}

The thermodynamic endpoint fixes one particular set of multipliers, but the exponential normalization is more general. Let
\begin{equation}
\bm\vartheta=(\beta,\bm\alpha),
\quad
A(\bm\alpha)=\sum_r\alpha_rP_r,
\label{eq:supp-family-theta}
\end{equation}
and define
\begin{equation}
\bar\rho_{\bm\vartheta}
=
\frac{e^{-\beta H_f-A(\bm\alpha)}}{Z_f(\bm\vartheta)},
\quad
Z_f(\bm\vartheta)=\Tr e^{-\beta H_f-A(\bm\alpha)},
\label{eq:supp-family-state}
\end{equation}
together with the sequential normalization
\begin{equation}
Z_f^{\rm seq}(\bm\vartheta)
=
\Tr\!\left[e^{-\beta H_f}e^{-A(\bm\alpha)}\right].
\label{eq:supp-family-Zseq}
\end{equation}
For the same physical trajectory \((n,m,r)\), define
\begin{equation}
\sigma_{\bm\vartheta}
=
\beta_iW_{mn}
+
(\beta-\beta_i)E_m^f
+
\alpha_r
+
\ln\frac{Z_f^{\rm seq}(\bm\vartheta)}{Z_i}.
\label{eq:supp-family-sigma}
\end{equation}
Repeating the cancellation in Note~\ref{supp:note-targetB-derivation} gives
\begin{equation}
\left\langle e^{-\sigma_{\bm\vartheta}}\right\rangle=1
\label{eq:supp-family-IFT}
\end{equation}
for every fixed \(\bm\vartheta\). The trajectory law is unchanged across the family; only the exponential reference used to weight each trajectory changes.

The physical thermodynamic member is
\begin{equation}
\bm\vartheta_{\rm th}=(\beta_f^\star,\bm\alpha_f),
\label{eq:supp-theta-th}
\end{equation}
because its maximum-entropy state reproduces the undisturbed endpoint energy and record. Ordinary Jarzynski sampling corresponds to
\begin{equation}
\bm\vartheta_J=(\beta_i,\bm 0),
\label{eq:supp-theta-J}
\end{equation}
for which \(\sigma_{\bm\vartheta_J}=\beta_i(W-\Delta F)\).

The revised sampling problem keeps the Jarzynski free-energy ratio fixed as the single physical target,
\begin{equation}
R_J
\equiv
e^{-\beta_i\Delta F}
=
\left\langle e^{-\beta_iW}\right\rangle.
\label{eq:supp-RJ}
\end{equation}
For any exact family member define
\begin{equation}
Y_{\bm\vartheta}=e^{-\sigma_{\bm\vartheta}},
\quad
C_{\bm\vartheta}=Y_{\bm\vartheta}-1.
\label{eq:supp-control-variable}
\end{equation}
Equation~\eqref{eq:supp-family-IFT} implies
\begin{equation}
\langle C_{\bm\vartheta}\rangle=0.
\label{eq:supp-control-zero}
\end{equation}
Hence for any fixed scalar \(c\),
\begin{equation}
\widehat R_J(\bm\vartheta,c)
=
\frac{1}{N}\sum_{j=1}^N
\left[
e^{-\beta_iW_j}-cC_{\bm\vartheta,j}
\right]
\label{eq:supp-controlled-estimator}
\end{equation}
is exactly unbiased,
\begin{equation}
\mathbb E\!\left[\widehat R_J(\bm\vartheta,c)\right]=R_J.
\label{eq:supp-controlled-unbiased}
\end{equation}
Thus the family is used here as a collection of exact control variables for one common free-energy target, not as a collection of different quantities to be estimated.

If \(\widehat{\bm\vartheta}\) and \(\widehat c\) are determined from an independent calibration or pilot sample and frozen before the production trajectories, Eq.~\eqref{eq:supp-controlled-unbiased} remains exact conditionally on those fitted values. Using the same finite sample both to choose the control and to evaluate Eq.~\eqref{eq:supp-controlled-estimator} would instead introduce statistical dependence and is not the protocol analyzed here.

The normalization \(Z_f^{\rm seq}(\bm\vartheta)/Z_i\) entering \(Y_{\bm\vartheta}\) must also be available independently of the production average. In the finite Bose--Hubbard benchmark it is evaluated from the calibrated endpoint model. The repetition counts below therefore quantify the fluctuation-run cost conditional on that calibration; they are not an end-to-end experimental resource count.

\suppsubsection{Thermodynamic displacement}

To measure how far an auxiliary reference moves from the actual endpoint, fix the additive gauge of the record multipliers and collect the independent retained operators in \(\bm F\). Define
\begin{equation}
\Phi(\bm\vartheta)=\ln Z_f(\bm\vartheta),
\quad
\bm m(\bm\vartheta)=\Tr(\bar\rho_{\bm\vartheta}\bm F).
\label{eq:supp-Phi-moments}
\end{equation}
Then \(\nabla_{\bm\vartheta}\Phi=-\bm m\), and the relative entropy from the physical thermodynamic member to another family member is
\begin{equation}
\mathcal D_{\rm th}(\bm\vartheta)
=
D(\bar\rho_{\rm th}\Vert\bar\rho_{\bm\vartheta})
=
\Phi(\bm\vartheta)-\Phi(\bm\vartheta_{\rm th})
+
(\bm\vartheta-\bm\vartheta_{\rm th})\cdot\bm m_{\rm th}.
\label{eq:supp-Dth}
\end{equation}
The thermodynamic member is the matching point at which the retained moments agree with the measured endpoint. Close to that point,
\begin{equation}
\mathcal D_{\rm th}
=
\frac{1}{2}
\Delta\bm\vartheta^{\mathsf T}
G_{\rm th}
\Delta\bm\vartheta
+
O(\|\Delta\bm\vartheta\|^3),
\label{eq:supp-local-D}
\end{equation}
where
\begin{equation}
G_{\rm th}
=
\nabla^2_{\bm\vartheta}\Phi\big|_{\bm\vartheta_{\rm th}}
\label{eq:supp-BKM}
\end{equation}
is the Bogoliubov--Kubo--Mori susceptibility matrix of the retained observables~\cite{Petz1994,LesniewskiRuskai1999,Scandi2019,Abiuso2020}. Pinsker's inequality implies that the trace-distance change of the reference state is at most \(\sqrt{\delta/2}\), and therefore bounds the change of any projective record in total variation by the same quantity.

\suppsubsection{Finite-confidence benchmark for the controlled free-energy estimator}

The numerical model has finite trajectory support, so the complete probability law of the estimator can be evaluated exactly. The exact model value \(R_J\) is used only to express the confidence event as a relative error and to benchmark different controls; the raw estimator in Eq.~\eqref{eq:supp-controlled-estimator} does not use the unknown \(\Delta F\).
Write
\begin{equation}
Y_J=\frac{e^{-\beta_iW}}{R_J},
\quad
\lambda=\frac{c}{R_J},
\label{eq:supp-normalized-controls}
\end{equation}
so that the normalized controlled variable is
\begin{equation}
G_{\bm\vartheta,\lambda}
=
Y_J
-
\lambda\left(Y_{\bm\vartheta}-1\right),
\quad
\langle G_{\bm\vartheta,\lambda}\rangle=1.
\label{eq:supp-G-control}
\end{equation}
For relative tolerance \(\varepsilon\), define the upper and lower Cram\'er rates
\begin{equation}
I_+
=
\sup_{t>0}
\left\{
t(1+\varepsilon)-K(t)
\right\},
\quad
I_-
=
\sup_{t<0}
\left\{
t(1-\varepsilon)-K(t)
\right\},
\label{eq:supp-cramer-rates}
\end{equation}
with
\begin{equation}
K(t)=\ln\left\langle e^{tG_{\bm\vartheta,\lambda}}\right\rangle.
\label{eq:supp-cgf}
\end{equation}
The two-sided Chernoff bound gives the sufficient model-conditioned repetition benchmark
\begin{equation}
N_{\Delta F}(\bm\vartheta,\lambda)
=
\frac{\ln(2/\eta)}{\min(I_+,I_-)}.
\label{eq:supp-N-control}
\end{equation}
We use \(\varepsilon=0.01\) and \(\eta=0.01\). This is a finite-support Cram\'er--Chernoff benchmark, not a claim of the mathematically minimal sample complexity.
For fixed \(\bm\vartheta\), the variance-minimizing coefficient is
\begin{equation}
\lambda_{\rm var}
=
\frac{\operatorname{Cov}(Y_J,Y_{\bm\vartheta})}
{\operatorname{Var}(Y_{\bm\vartheta})},
\label{eq:supp-lambda-var}
\end{equation}
and the corresponding variance is reduced by the usual correlation factor. The finite-confidence optimum need not equal \(\lambda_{\rm var}\), because Eq.~\eqref{eq:supp-N-control} depends on the complete tail law rather than only the second moment. In the reported benchmark, \(\lambda\) is optimized directly for Eq.~\eqref{eq:supp-N-control}. Experimentally, the dimensional coefficient \(c\) can be estimated on an independent pilot sample and then frozen; that pilot cost is separate from the production-run benchmark below.

\suppsubsection{Common-target numerical comparison}

For the nominal Bose--Hubbard endpoint, the ordinary Jarzynski free-energy difference is
\begin{equation}
\Delta F/J=-10.291645531969.
\label{eq:supp-DeltaF-value}
\end{equation}
The direct Jarzynski estimator has the benchmark
\begin{equation}
N_J=5.64630461\times10^9.
\label{eq:supp-NJ-new}
\end{equation}
At exact endpoint matching, \(\bm\vartheta=\bm\vartheta_{\rm th}\), the optimized control coefficient gives
\begin{equation}
N_{\rm th}=3.58726882\times10^9,
\label{eq:supp-Nth-new}
\end{equation}
a \(36.467\%\) reduction relative to direct Jarzynski sampling. The correlation between the normalized Jarzynski weight and the thermodynamic exact weight is
\begin{equation}
\operatorname{Corr}(Y_J,Y_{\rm th})=0.6216467.
\label{eq:supp-corr-th}
\end{equation}
Thus the physical endpoint member already carries statistically useful information about the rare work fluctuations that dominate the Jarzynski estimate.

At the common information budget \(\delta=0.01\), variance optimization over the admissible reference and control coefficient gives
\begin{equation}
N_{\rm var}=3.25004459\times10^9,
\label{eq:supp-Nvar-new}
\end{equation}
whereas direct minimization of the two-sided finite-confidence benchmark gives
\begin{equation}
N_{\rm FC}=2.94955403\times10^9.
\label{eq:supp-Nfc-new}
\end{equation}
The latter corresponds to a \(47.761\%\) reduction relative to direct Jarzynski sampling and a \(17.777\%\) reduction relative to the endpoint-matching thermodynamic control. The controlled estimator variance is not smallest at this finite-confidence optimum, confirming that the tail-sensitive and second-moment objectives are distinct.

\begin{figure}[t]
\includegraphics[width=0.8\textwidth]{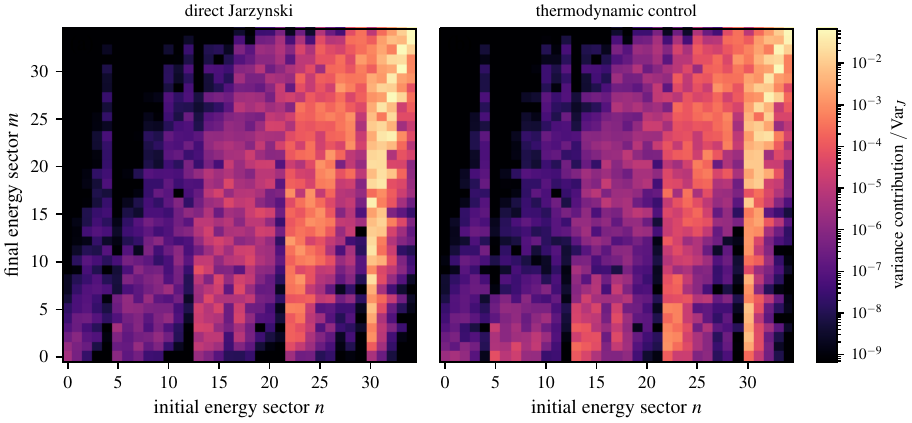}
\caption{
Trajectory-resolved variance burden for the common free-energy target.
(a) Contribution of each initial--final energy-sector pair to the direct Jarzynski estimator variance, normalized by the total direct-Jarzynski variance.
(b) The corresponding contribution after applying the endpoint-matching thermodynamic control to the same trajectory law.
The control suppresses part of the rare-sector burden without altering the sampled trajectories.
}
\label{fig:supp-deltaF-variance}
\end{figure}

Figure~\ref{fig:supp-deltaF-variance} resolves where the variance reduction occurs in the \((n,m)\) trajectory plane.

\begin{figure}[t]
\includegraphics[width=\textwidth]{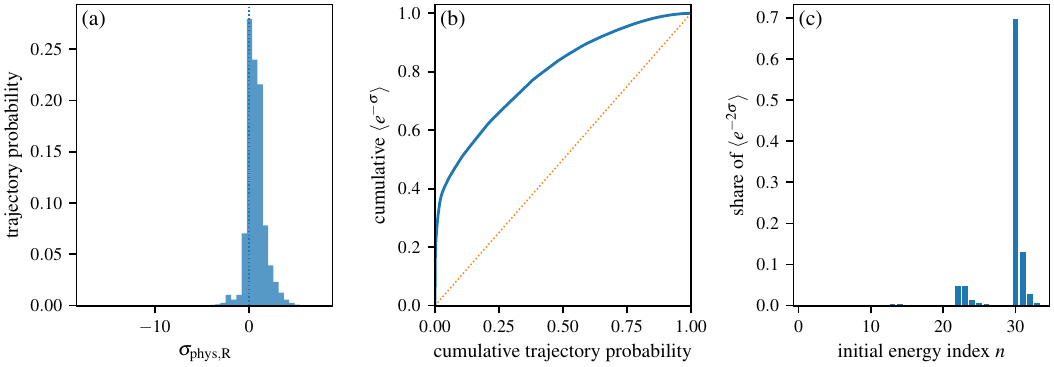}
\caption{
Rare-event structure of the endpoint-matching thermodynamic member.
(a) Probability distribution of \(\sigma_{\rm phys,R}\).
(b) Cumulative contribution to \(\langle e^{-\sigma_{\rm phys,R}}\rangle\) when trajectories are ordered from the most negative \(\sigma\) upward.
(c) Contribution of each initial energy sector to \(\langle e^{-2\sigma_{\rm phys,R}}\rangle\).
This figure characterizes the auxiliary thermodynamic weight itself; the free-energy control-variate performance is shown separately in Fig.~\ref{fig:supp-deltaF-variance}.
}
\label{fig:supp-rare}
\end{figure}

The original thermodynamic fluctuation variable itself remains strongly rare-event weighted. Figure~\ref{fig:supp-rare} shows its probability distribution, cumulative exponential contribution, and second-moment concentration. These diagnostics explain why the exact thermodynamic identity can be statistically difficult when used as a target by itself, even though it is useful as a correlated zero-mean control for Jarzynski.

\suppsubsection{Constrained information--sampling frontier and square-root onset}

We now optimize the controlled estimator while allowing only a prescribed displacement from the physical endpoint,
\begin{equation}
N_{\Delta F}^{\star}(\delta)
=
\min_{\bm\vartheta,\lambda}
N_{\Delta F}(\bm\vartheta,\lambda)
\quad
\text{subject to}
\quad
\mathcal D_{\rm th}(\bm\vartheta)\leq\delta.
\label{eq:supp-frontier-new}
\end{equation}
The numerical frontier is deliberately kept local to the thermodynamic endpoint; it does not approach the Jarzynski member itself, whose normalization is the unknown target. In the nominal model,
\begin{equation}
D(\bar\rho_{\rm th}\Vert\bar\rho_{\bm\vartheta_J})=0.5420881,
\label{eq:supp-D-J}
\end{equation}
whereas the reported frontier uses \(0\leq\delta\leq0.05\).

To obtain the local law, let
\begin{equation}
\ell(\bm\vartheta)
=
\ln N_{\Delta F}^{\star}(\bm\vartheta)
\label{eq:supp-logN}
\end{equation}
denote the finite-confidence log cost after profiling over \(\lambda\), and set
\begin{equation}
\bm g
=
\nabla_{\bm\vartheta}\ell\big|_{\bm\vartheta_{\rm th}}.
\label{eq:supp-g-new}
\end{equation}
Combining the linear expansion of \(\ell\) with Eq.~\eqref{eq:supp-local-D}, the leading optimal displacement is
\begin{equation}
\Delta\bm\vartheta^\star
=
-\sqrt{
\frac{2\delta}{\bm g^{\mathsf T}G_{\rm th}^{-1}\bm g}
}
G_{\rm th}^{-1}\bm g
+
O(\delta).
\label{eq:supp-natural-direction-new}
\end{equation}
Therefore
\begin{equation}
\ln\frac{N_{\Delta F}^{\star}(\delta)}{N_{\rm th}}
=
-\sqrt{2\delta\,\bm g^{\mathsf T}G_{\rm th}^{-1}\bm g}
+
O(\delta),
\label{eq:supp-logN-sqrt}
\end{equation}
and hence
\begin{equation}
1-
\frac{N_{\Delta F}^{\star}(\delta)}{N_{\rm th}}
=
C_{\Delta F}\sqrt\delta+O(\delta)
\label{eq:supp-N-sqrt-new}
\end{equation}
with
\begin{equation}
C_{\Delta F}
=
\sqrt{2\bm g^{\mathsf T}G_{\rm th}^{-1}\bm g}
\simeq1.69178.
\label{eq:supp-C-DeltaF}
\end{equation}
The square root has the same geometric origin as before: thermodynamic displacement is quadratic in \(\Delta\bm\vartheta\), whereas the first statistical gain is linear. What changes is the objective: the optimized quantity is now the sampling cost of one fixed physical target, \(\Delta F\).

\begin{figure}[t]
\includegraphics[width=\textwidth]{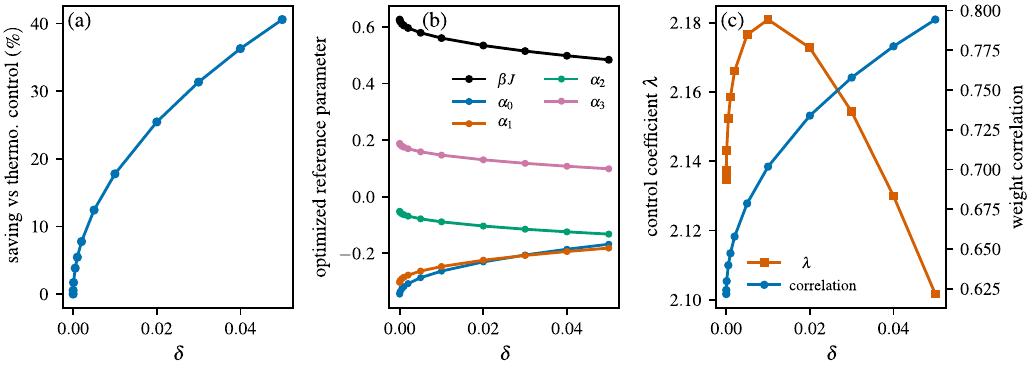}
\caption{
Detailed common-target information--sampling frontier.
(a) Percentage reduction of the best-found \(\Delta F\) sampling benchmark relative to the endpoint-matching thermodynamic control.
(b) Gauge-fixed auxiliary-reference parameters \(\beta J\) and \(\alpha_0,\ldots,\alpha_3\) along the frontier.
(c) Profiled control coefficient \(\lambda\) and correlation between the Jarzynski and auxiliary exact weights.
All points estimate the same Jarzynski free-energy ratio; only the exact control reference is displaced.
}
\label{fig:supp-deltaF-frontier}
\end{figure}

Figure~\ref{fig:supp-deltaF-frontier} gives the detailed path of the optimized auxiliary reference.

At \(\delta=0.01\), Eq.~\eqref{eq:supp-Nfc-new} gives the main-text operating point. Over the displayed local range the best-found cost continues to decrease; at the largest reported displacement, \(\delta\simeq0.05\), the benchmark is \(0.5918N_{\rm th}\). We do not interpret this monotonic numerical continuation as a global optimum beyond the displayed local neighborhood.

\suppsubsection{What the reported resource count includes}

The repetition numbers above count production TPM runs after the endpoint reference and control coefficient have been fixed. For the canonical preparation, a complete experimental accounting has the schematic form
\begin{equation}
N_{\rm tot}
=
N_{{\rm cal},f}
+
N_{\rm pilot}
+
N_{\rm fluc},
\label{eq:supp-total-cost-canonical}
\end{equation}
where \(N_{{\rm cal},f}\) determines the undisturbed endpoint energy and record and hence \(\bm\vartheta_{\rm th}\), \(N_{\rm pilot}\) fixes the control coefficient without using the production sample, and \(N_{\rm fluc}\) is the Cram\'er--Chernoff benchmark analyzed here. For arbitrary initial states an additional initial-state calibration cost appears, as discussed in Note~\ref{supp:note-arbitrary-initial}.

The present frontier isolates \(N_{\rm fluc}\) in order to answer a specific question: after physically meaningful endpoint information has been calibrated, how much can that information reduce the rare-event cost of estimating the same \(\Delta F\) from the same trajectory law? A complete laboratory optimization must minimize the total resource count in Eq.~\eqref{eq:supp-total-cost-canonical} and include uncertainty in the calibrated normalization of the auxiliary exact relation.

\suppnote{Arbitrary initial quantum states}
\label{supp:note-arbitrary-initial}

The main derivation used a canonical initial state, for which the first TPM measurement leaves the ensemble unchanged. We now remove that assumption and identify the additional information required by the measurement-defined process.

\suppsubsection{What the first TPM measurement actually sees}

For a general microscopic preparation \(\rho_i\), the first TPM energy measurement need not leave the ensemble state unchanged.
Its nonselective action produces
\begin{equation}
\rho_i^{\rm d}
=
\cD_{H_i}[\rho_i]
=
\sum_n
\Pi_n^i
\rho_i
\Pi_n^i.
\label{eq:supp-rhoid}
\end{equation}
The state that subsequently evolves in the TPM protocol is therefore
\begin{equation}
\rho_f^{\rm TPM}
=
U
\rho_i^{\rm d}
U^\dagger.
\label{eq:supp-rhof-TPM}
\end{equation}
This should be distinguished from the unmeasured coherent endpoint
\(U\rho_iU^\dagger\).
The arbitrary-state fluctuation relation below describes the explicitly measurement-defined TPM process.

To isolate the departure of this dephased state from a canonical reference, choose an arbitrary full-rank Gibbs state for the initial Hamiltonian,
\begin{equation}
\tau_i^{(0)}
=
\frac{
e^{-\beta_0H_i}
}{
Z_i^{(0)}
},
\quad
Z_i^{(0)}
=
\Tr
e^{-\beta_0H_i}.
\label{eq:supp-reference-initial}
\end{equation}
For full-rank \(\rho_i^{\rm d}\), define the initial-information operator
\begin{equation}
\Gamma_i
=
\ln
\rho_i^{\rm d}
-
\ln
\tau_i^{(0)}.
\label{eq:supp-Gamma}
\end{equation}
Because \(\rho_i^{\rm d}\) is block diagonal in the eigenspaces of \(H_i\),
\(\Gamma_i\) commutes with \(H_i\).
Physically, \(\Gamma_i\) is an initial-state correction dictionary: in each energy-resolved sector it records the logarithmic difference between the actual dephased preparation and the chosen Gibbs reference.

Within each energy block, diagonalize \(\rho_i^{\rm d}\),
\begin{equation}
\rho_i^{\rm d}
=
\sum_{n,a}
\lambda_{na}
R_{na},
\quad
R_{na}
\leq
\Pi_n^i.
\label{eq:supp-rhoid-spectral}
\end{equation}
The projectors \(R_{na}\) jointly refine the initial energy and the information operator.
Define
\begin{equation}
q_n^{(0)}
=
\frac{
e^{-\beta_0E_n^i}
}{
Z_i^{(0)}
},
\quad
\gamma_{na}
=
\ln
\frac{
\lambda_{na}
}{
q_n^{(0)}
}.
\label{eq:supp-gamma}
\end{equation}
The number \(\gamma_{na}\) is the trajectory-level value of \(\Gamma_i\).

The refined trajectory probability is
\begin{equation}
p(n,a,m,r)
=
\lambda_{na}
\Tr
\left[
P_r
\Pi_m^f
U
R_{na}
U^\dagger
\Pi_m^f
P_r
\right].
\label{eq:supp-arb-joint}
\end{equation}
If \(R_{na}\) has rank larger than one, the trace automatically includes that multiplicity.

\suppsubsection{Exact arbitrary-state relation}

Let
\begin{equation}
\Psi_i^{(0)}
=
-\ln
Z_i^{(0)},
\quad
\Delta\Psi^{(0)}
=
\Psi_f-\Psi_i^{(0)}.
\label{eq:supp-arb-Psi}
\end{equation}
The arbitrary-state trajectory variable is
\begin{equation}
\begin{aligned}
\sigma_{\rm arb}
=
\beta_0W_{mn}
+
\left(
\beta_f^\star-\beta_0
\right)
E_m^f
+
\alpha_{r,f}
+
\gamma_{na}
+
\mathcal Q_{\rm nc}
-
\Delta\Psi^{(0)}.
\end{aligned}
\label{eq:supp-sigma-arb}
\end{equation}
For a full-rank initial dephased state,
\begin{equation}
\left\langle
e^{-\sigma_{\rm arb}}
\right\rangle
=
1.
\label{eq:supp-arb-IFT}
\end{equation}

Although Eq.~\eqref{eq:supp-sigma-arb} is written using a Gibbs reference, the choice of \(\beta_0\) is only bookkeeping.
Substituting Eqs.~\eqref{eq:supp-work} and \eqref{eq:supp-gamma} into Eq.~\eqref{eq:supp-sigma-arb} cancels it exactly,
\begin{equation}
\sigma_{\rm arb}
=
\beta_f^\star E_m^f
+
\alpha_{r,f}
+
\ln
\lambda_{na}
+
\mathcal Q_{\rm nc}
-
\Psi_f.
\label{eq:supp-arb-reference-independent}
\end{equation}
Thus different choices of the auxiliary Gibbs reference give the same complete trajectory variable.

The average initial correction has a direct information meaning,
\begin{equation}
\left\langle
\gamma
\right\rangle
=
D
\left(
\rho_i^{\rm d}
\Vert
\tau_i^{(0)}
\right)
\geq
0.
\label{eq:supp-mean-gamma}
\end{equation}
At the final endpoint, the same steps used in Note~\ref{supp:note-targetB-derivation} give
\begin{equation}
\left\langle
\sigma_{\rm arb}
\right\rangle
=
D
\left(
\rho_f^{\rm TPM}
\Vert
\bar\rho_f
\right)
+
\Delta_{\rm meas}
+
\mathcal Q_{\rm nc}.
\label{eq:supp-arb-mean}
\end{equation}
Here \(\bar\rho_f\) is fitted to the state in Eq.~\eqref{eq:supp-rhof-TPM} before the final energy measurement is performed.
Thus the endpoint representative is again calibrated from an undisturbed copy of the process whose fluctuation statistics are later measured.

For a nondegenerate initial Hamiltonian,
\begin{equation}
\rho_i^{\rm d}
=
\sum_n
p_n
\left|
E_n^i
\right\rangle
\left\langle
E_n^i
\right|,
\label{eq:supp-nondegenerate-rho}
\end{equation}
and the correction reduces to
\begin{equation}
\gamma_n
=
\ln
\frac{
p_n
}{
q_n^{(0)}
}.
\label{eq:supp-gamma-simple}
\end{equation}
Once the probabilities \(p_n\) have been calibrated, the ordinary initial energy outcome \(n\) already identifies \(\gamma_n\).
No extra trajectory measurement is then needed.
For a degenerate \(H_i\), a finer label \(a\) may be required.

\suppsubsection{Calibration closure versus trajectory closure}

The arbitrary-state theorem creates two separate operational questions: whether the retained preparation data determine the required correction, and whether each fluctuation run reveals the corresponding trajectory value.

First, the retained initial macro-record must contain enough information to determine \(\Gamma_i\).
We call this \emph{calibration closure}.
For a preparation class
\(\mathcal C_{\mathcal R_i}\)
compatible with one retained initial record, full-rank closure is equivalent to
\begin{equation}
\cD_{H_i}[\rho]
=
\cD_{H_i}[\sigma]
\quad
\text{for all }
\rho,\sigma
\in
\mathcal C_{\mathcal R_i}.
\label{eq:supp-closure-rho}
\end{equation}
Indeed, because the matrix logarithm is one-to-one on positive operators,
Eq.~\eqref{eq:supp-closure-rho} is equivalent to
\begin{equation}
\Gamma_i[\rho]
=
\Gamma_i[\sigma]
\quad
\text{for all }
\rho,\sigma
\in
\mathcal C_{\mathcal R_i}.
\label{eq:supp-closure-Gamma}
\end{equation}

Second, once \(\Gamma_i\) is known, the fluctuation experiment must resolve the corresponding realization-level value \(\gamma\).
We call this \emph{trajectory measurement closure}.
The two requirements are not the same.
Knowing the expectation value of \(\Gamma_i\), or even knowing that \(\Gamma_i\) lies in the span of some coarse operators, does not by itself guarantee that the trajectory measurement reveals its eigenvalue.

For a nondegenerate \(H_i\), the situation is simple.
The initial energy outcome \(n\) already labels the diagonal value \(\gamma_n\).
For a degenerate \(H_i\), one may need the common spectral refinement \(R_{na}\) in Eq.~\eqref{eq:supp-rhoid-spectral}.

An approximate version is also useful.
If every compatible dephased state obeys
\begin{equation}
\rho_i^{\rm d}
\geq
\mu I
\label{eq:supp-mu}
\end{equation}
for some \(\mu>0\), then
\begin{equation}
\left\|
\Gamma_i[\rho]
-
\Gamma_i[\sigma]
\right\|_\infty
\leq
\frac{1}{\mu}
\left\|
\cD_{H_i}
\left[
\rho-\sigma
\right]
\right\|_\infty.
\label{eq:supp-approx-closure}
\end{equation}
The bound becomes loose when very small populations are allowed, which already suggests that rare initial energy sectors can make calibration expensive.

\suppsubsection{Why the natural \((E,R)\) record is not enough for unrestricted preparations}

The nominal initial Hamiltonian is nondegenerate; its smallest resolved level spacing is
\begin{equation}
\Delta E_{\rm min}
=
3.50\times10^{-4}J.
\label{eq:supp-min-gap}
\end{equation}
An arbitrary energy-dephased preparation is therefore specified by \(35\) populations \(p_n\), or \(34\) independent numbers after normalization.

For an energy-diagonal state, define
\begin{equation}
R_{rn}
=
\left\langle
E_n^i
\right|
P_r
\left|
E_n^i
\right\rangle.
\label{eq:supp-Rrn}
\end{equation}
The constraints supplied by normalization, mean energy, and the spatial record are linear in the population vector.
Their constraint matrix is
\begin{equation}
C
=
\begin{pmatrix}
1\\
E_n^i\\
R_{0n}\\
R_{1n}\\
R_{2n}\\
R_{3n}\\
R_{4n}
\end{pmatrix}.
\label{eq:supp-Cmatrix}
\end{equation}
Because the five record probabilities sum to one, one row is dependent.
Numerically,
\begin{equation}
\operatorname{rank}
C
=
6.
\label{eq:supp-rankC}
\end{equation}
The compatible energy-population fiber therefore has dimension
\begin{equation}
35-6
=
29.
\label{eq:supp-fiber-dimension}
\end{equation}
Thus the initial \((E,R)\) macro-record does not determine the energy populations and hence does not determine \(\Gamma_i\) over an unrestricted preparation class.

To make this failure explicit, we construct two full-rank energy-diagonal states with exactly the same initial mean energy and spatial record.
We first maximize and minimize the second energy moment over the compatible population polytope, and then mix each extreme distribution with the canonical distribution,
\begin{equation}
\mathbf p^{(A)}
=
0.8
\mathbf p_{\rm can}
+
0.2
\mathbf p_{\rm ext}^{(+)},
\quad
\mathbf p^{(B)}
=
0.8
\mathbf p_{\rm can}
+
0.2
\mathbf p_{\rm ext}^{(-)}.
\label{eq:supp-pair-construction}
\end{equation}
The common \(80\%\) canonical part keeps every population strictly positive while preserving all linear constraints.

The two preparations satisfy
\begin{equation}
D_{\rm TV}
\left(
\mathbf p^{(A)},
\mathbf p^{(B)}
\right)
=
0.200000,
\label{eq:supp-pair-TV}
\end{equation}
and
\begin{equation}
\left\|
\Gamma_i^{(A)}
-
\Gamma_i^{(B)}
\right\|_\infty
=
1.85551.
\label{eq:supp-pair-Gamma}
\end{equation}
After the same work stroke,
\begin{equation}
\left|
E_f^{(A)}
-
E_f^{(B)}
\right|
=
0.121411J,
\label{eq:supp-pair-Ef}
\end{equation}
and their final spatial records differ by
\begin{equation}
D_{\rm TV}
\left(
\mathbf p_{R,f}^{(A)},
\mathbf p_{R,f}^{(B)}
\right)
=
0.0320344.
\label{eq:supp-pair-final-TV}
\end{equation}

The connection to the return decomposition can be made exact for this pair.
Because both preparations are energy diagonal, the first TPM measurement leaves them unchanged.
Because they also have the same initial mean energy and spatial record, they share the same maximum-entropy representative \(\bar\rho_{E,R,i}\).
We may therefore write
\begin{equation}
\rho_i^{(X)}
=
\bar\rho_{E,R,i}
+
\chi_i^{(X)},
\quad
X\in\{A,B\},
\label{eq:supp-compatible-chi}
\end{equation}
with different hidden parts \(\chi_i^{(A)}\neq\chi_i^{(B)}\).
Their microscopic difference lies entirely in this unresolved sector,
\begin{equation}
\rho_i^{(A)}-\rho_i^{(B)}
=
\chi_i^{(A)}-\chi_i^{(B)}.
\label{eq:supp-compatible-hidden-difference}
\end{equation}
After the same unitary work stroke, the difference of the final spatial records is therefore
\begin{equation}
p_{s,f}^{(A)}-p_{s,f}^{(B)}
=
\Tr\!\left[
P_sU\left(\chi_i^{(A)}-\chi_i^{(B)}\right)U^\dagger
\right]
=
r_s^{(A)}-r_s^{(B)}.
\label{eq:supp-compatible-return-record}
\end{equation}
Likewise, their final mean-energy difference is
\begin{equation}
E_f^{(A)}-E_f^{(B)}
=
\Tr\!\left[
H_fU\left(\chi_i^{(A)}-\chi_i^{(B)}\right)U^\dagger
\right]
=
r_E^{(A)}-r_E^{(B)}.
\label{eq:supp-compatible-return-energy}
\end{equation}
Here \(r_s^{(X)}\) and \(r_E^{(X)}\) are the record and energy return contributions defined in Eqs.~\eqref{eq:supp-ER-record-return-component} and \eqref{eq:supp-ER-energy-return-components} for preparation \(X\).
Thus the same microscopic freedom that prevents the initial macro-record from fixing \(\Gamma_i\) is dynamically active: it also produces different later accessible records under the same control protocol.

\begin{figure}[t]
\includegraphics[width=\textwidth]{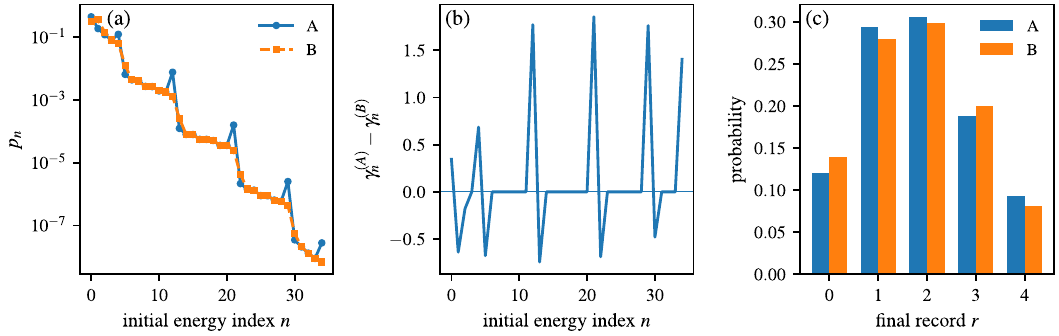}
\caption{
Explicit failure of initial \((E,R)\) closure for unrestricted preparations.
The two states have exactly the same initial mean energy and spatial record.
(a) Their energy populations \(p_n^{(A)}\) and \(p_n^{(B)}\) are nevertheless different.
(b) The difference of the resulting trajectory corrections, \(\gamma_n^{(A)}-\gamma_n^{(B)}\).
(c) Final spatial records after applying the same unitary work protocol.
The later difference shows that the hidden initial energy populations are dynamically active, not merely redundant microscopic detail.
}
\label{fig:supp-closure}
\end{figure}

Figure~\figpanel{fig:supp-closure}{a} shows the unresolved difference already present inside the same initial macro-record.
Figure~\figpanel{fig:supp-closure}{b} shows that this hidden difference changes the correction required by the arbitrary-state fluctuation theorem, while Fig.~\figpanel{fig:supp-closure}{c} shows its later return into the accessible spatial record.
Equations~\eqref{eq:supp-compatible-return-record} and \eqref{eq:supp-compatible-return-energy} make the connection to Ref.~\cite{Ahmadi2026Endogenous} explicit: information absent from the present thermodynamic state is carried by the unresolved component and can reappear after microscopic evolution.
The failure of closure therefore has nothing to do with initial coherence; both preparations are already diagonal in energy.

The exact theorem was also checked numerically for several full-rank preparations. Besides the canonical state and a second Gibbs state, we use an independent non-Gibbs test preparation and the same-\((E,R)\) pair constructed above.

\begin{table}[t]
\caption{
Full-rank arbitrary-initial-state checks using the DOP853 propagation.
Each preparation is fitted to its own TPM-defined endpoint.
The last column verifies the corresponding exact IFT.
}
\label{tab:supp-arbitrary-states}
\begin{ruledtabular}
\begin{tabular}{lrrrrrrr}
Preparation
&
\(E_i/J\)
&
\(E_f/J\)
&
\(D\)
&
\(\Delta_{\rm meas}\)
&
\(\mathcal Q_{\rm nc}\)
&
\(\langle\sigma\rangle\)
&
IFT
\\
\hline
Canonical
&
\(0.571302\)
&
\(-3.801855\)
&
\(0.641610\)
&
\(0.0313255\)
&
\(0.00187891\)
&
\(0.674814\)
&
\(1.000000000000005\)
\\
Gibbs, \(\beta J=0.12\)
&
\(3.810731\)
&
\(-3.120386\)
&
\(0.534103\)
&
\(0.0130652\)
&
\(0.000710290\)
&
\(0.547879\)
&
\(1.000000000000003\)
\\
Non-Gibbs mixture
&
\(2.910576\)
&
\(-3.395082\)
&
\(0.517145\)
&
\(0.0214458\)
&
\(0.00102770\)
&
\(0.539618\)
&
\(0.999999999999999\)
\\
Compatible A
&
\(0.571302\)
&
\(-3.721259\)
&
\(0.724064\)
&
\(0.0310373\)
&
\(0.00203466\)
&
\(0.757136\)
&
\(0.999999999999999\)
\\
Compatible B
&
\(0.571302\)
&
\(-3.842670\)
&
\(0.657424\)
&
\(0.0301950\)
&
\(0.00136715\)
&
\(0.688986\)
&
\(1.000000000000000\)
\\
\end{tabular}
\end{ruledtabular}
\end{table}

The different values in Table~\ref{tab:supp-arbitrary-states} show that the theorem is not tied to a Gibbs preparation.
The IFT remains normalized for each full-rank preparation when the correct initial information \(\Gamma_i\) is used.

\suppsubsection{A coherent preparation with canonical energy populations}

For the nondegenerate nominal initial Hamiltonian, the TPM construction can also be insensitive to large amounts of initial coherence.
Consider
\begin{equation}
\left|
\psi_i
\right\rangle
=
\sum_n
\sqrt{
q_n
}
e^{0.2in^2}
\left|
E_n^i
\right\rangle,
\label{eq:supp-coherent-state}
\end{equation}
where \(q_n\) are the canonical energy populations.
Then
\begin{equation}
\cD_{H_i}
\left[
\left|
\psi_i
\right\rangle
\left\langle
\psi_i
\right|
\right]
=
\rho_i^{\rm can},
\label{eq:supp-coherent-dephase}
\end{equation}
so
\begin{equation}
\Gamma_i
=
0.
\label{eq:supp-coherent-Gamma}
\end{equation}
The complete TPM statistics are therefore the same as for the canonical mixed state.

This equality of TPM statistics does not mean that the coherence is dynamically irrelevant.
For the same unitary protocol, the unmeasured coherent endpoint and the TPM endpoint have trace distance
\begin{equation}
T
=
0.762029,
\label{eq:supp-coherent-trace}
\end{equation}
and their final spatial records differ by
\begin{equation}
D_{\rm TV}
=
0.249845.
\label{eq:supp-coherent-record}
\end{equation}
The TPM theorem is blind to this difference because the first energy measurement removes the coherence before the work protocol begins.
Viewed through the same unresolved-state logic, the coherence is microscopic information that the initial energy measurement does not retain; under the unmeasured unitary it can later become visible in the spatial record, but TPM deliberately deletes it before that return can occur.
This example is therefore a reminder that the arbitrary-state result is a theorem for the measurement-defined process, not for the corresponding unmeasured coherent evolution.

\suppsubsection{Rank-deficient preparations and support deficit}

If \(\rho_i^{\rm d}\) is rank deficient, \(\ln\rho_i^{\rm d}\) is not finite outside its support.
Let
\begin{equation}
S_i
=
\operatorname{supp}
\rho_i^{\rm d}
\label{eq:supp-support-projector}
\end{equation}
be the projector onto the populated initial subspace. The trajectory construction remains well defined on that support.

Define the positive endpoint operator
\begin{equation}
B_f^{\rm seq}
=
\sum_m
e^{-\beta_f^\star E_m^f}
\Pi_m^f
e^{-A_f}
\Pi_m^f.
\label{eq:supp-Bseq}
\end{equation}
Its trace is
\begin{equation}
\Tr
B_f^{\rm seq}
=
Z_f^{\rm seq}.
\label{eq:supp-Bseq-trace}
\end{equation}
The exponential average now becomes
\begin{equation}
\left\langle
e^{-\sigma_{\rm arb}}
\right\rangle
=
\frac{
\Tr
\left[
B_f^{\rm seq}
U
S_i
U^\dagger
\right]
}{
Z_f^{\rm seq}
}
=
1-\Lambda_{\rm supp},
\label{eq:supp-support-IFT}
\end{equation}
where
\begin{equation}
\Lambda_{\rm supp}
=
1
-
\frac{
\Tr
\left[
B_f^{\rm seq}
U
S_i
U^\dagger
\right]
}{
Z_f^{\rm seq}
}.
\label{eq:supp-Lambda-supp}
\end{equation}
Since
\(0\leq US_iU^\dagger\leq I\)
and \(B_f^{\rm seq}\geq0\),
\begin{equation}
0
\leq
\Lambda_{\rm supp}
\leq
1.
\label{eq:supp-Lambda-bounds}
\end{equation}
We call this quantity the \emph{support deficit}.
It measures the fraction of the sequential endpoint weight that cannot be reached from the populated initial support. It is structurally related to support-based corrections in fluctuation theorems, but we do not identify it with absolute irreversibility because no explicit reverse path measure is constructed here.

\begin{figure}[t]
\includegraphics[width=\textwidth]{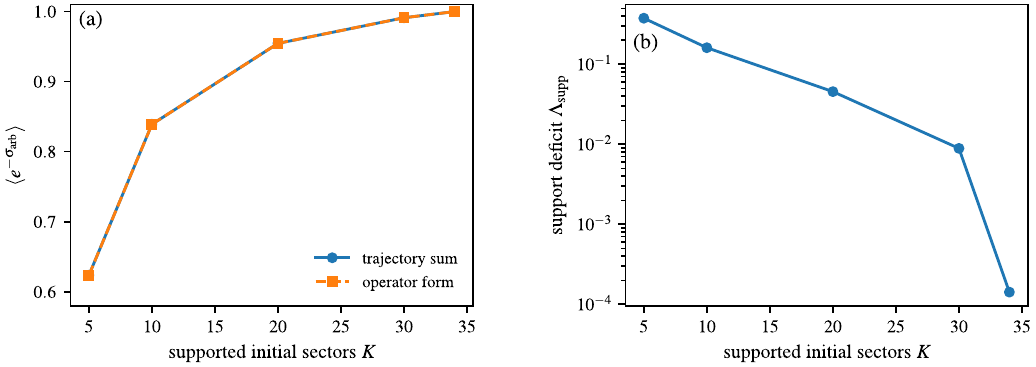}
\caption{
Rank-deficient initial preparations.
(a) Direct trajectory evaluation of \(\langle e^{-\sigma_{\rm arb}}\rangle\) and the operator expression in Eq.~\eqref{eq:supp-support-IFT} as the number \(K\) of supported initial energy sectors is increased.
The two calculations agree to numerical precision.
(b) Corresponding support deficit \(\Lambda_{\rm supp}\) on a logarithmic scale.
The deficit vanishes as the missing initial support is restored.
}
\label{fig:supp-support}
\end{figure}

We test Eq.~\eqref{eq:supp-support-IFT} by retaining only the lowest \(K\) canonical energy sectors and renormalizing the state. Figure~\ref{fig:supp-support} compares the direct trajectory sum with the operator expression.

Figure~\figpanel{fig:supp-support}{a} verifies the same quantity in two independent ways.
For \(K=5,10,20,30,34\), the exponential averages are respectively
\begin{equation}
0.623569,
\quad
0.839009,
\quad
0.954502,
\quad
0.991113,
\quad
0.999858.
\label{eq:supp-support-values}
\end{equation}
Figure~\figpanel{fig:supp-support}{b} shows the complementary deficits.
The largest DOP853--Magnus difference in these tests is
\begin{equation}
2.85\times10^{-11}.
\label{eq:supp-support-solver}
\end{equation}

\suppsubsection{Initial-state calibration cost}

For the nominal Bose--Hubbard model, whose initial and final Hamiltonians are nondegenerate in the working sector, the arbitrary-state correction only needs the initial energy populations \(p_n\). This is much less information than full state tomography, but the logarithm in
\(\gamma_n=\ln(p_n/q_n^{(0)})\) makes very rare populations difficult to estimate.

Suppose the true initial populations are \(p_n\), while an independent calibration ensemble produces estimates \(\widehat p_n\).
Define
\begin{equation}
T_{mn}
=
\left|
\left\langle
E_m^f
\right|
U
\left|
E_n^i
\right\rangle
\right|^2
\label{eq:supp-Tmn}
\end{equation}
and
\begin{equation}
R_{mr}
=
\left\langle
E_m^f
\right|
P_r
\left|
E_m^f
\right\rangle.
\label{eq:supp-Rmr}
\end{equation}
The normalized sensitivity of the exponential average to the initial sector \(n\) is
\begin{equation}
b_n
=
\frac{1}{Z_f^{\rm seq}}
\sum_{m,r}
T_{mn}
R_{mr}
e^{-\beta_f^\star E_m^f-\alpha_{r,f}},
\quad
\sum_n
b_n
=
1.
\label{eq:supp-bn}
\end{equation}
If \(\widehat p_n\) is used in place of the true \(p_n\), the measured exponential average is exactly
\begin{equation}
\mathcal J
\left(
\widehat{\mathbf p}
\right)
=
\sum_n
b_n
\frac{
p_n
}{
\widehat p_n
}.
\label{eq:supp-calibration-exact}
\end{equation}
Unlike the final exponential-family parameters in Note~\ref{supp:note-sampling-frontier}, an arbitrary error in the initial probabilities does not merely choose another exact member of the same theorem. The correction must match the actual initial energy distribution.

For a multinomial calibration using \(N_{\rm cal}\) independent copies, a delta-method expansion gives
\begin{equation}
\operatorname{Var}
\left[\mathcal J\right]\simeq
\frac{1}{N_{\rm cal}}\left(\sum_n \frac{b_n^2}{p_n} - 1\right),
\label{eq:supp-calibration-var}
\end{equation}
while the leading nonlinear bias is
\begin{equation}
\mathbb E
\left[
\mathcal J
\right]
-
1
\simeq
\frac{1}{N_{\rm cal}}
\sum_n
b_n
\frac{
1-p_n
}{
p_n
}.
\label{eq:supp-calibration-bias}
\end{equation}
Both expressions show why rare initial sectors matter: errors in a small \(p_n\) are amplified by the reciprocal probability.

\begin{figure}[t]
\includegraphics[width=0.72\textwidth]{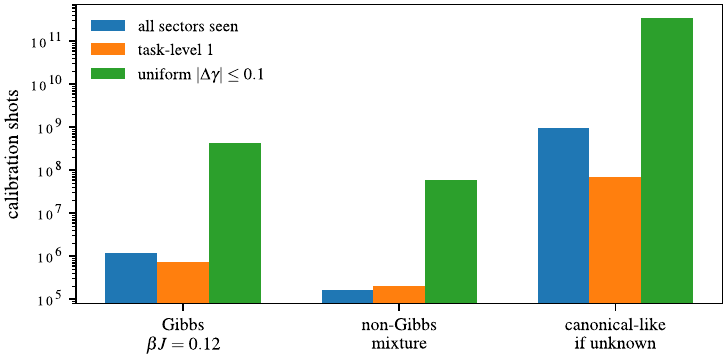}
\caption{
Initial-state calibration cost for three full-support preparations.
The vertical scale is logarithmic.
The three bars for each preparation show: a conservative \(99\%\) bound for observing all \(35\) energy sectors at least once; a local task-level estimate for controlling the fluctuation normalization to \(1\%\) at \(99\%\) confidence; and a conservative bound for estimating every trajectory correction with \(|\Delta\gamma_n|\leq0.1\) at \(99\%\) confidence.
The large separation shows that reconstructing every rare-sector correction can be much harder than controlling the fluctuation task itself.
}
\label{fig:supp-metrology}
\end{figure}

We evaluate these costs for the Gibbs preparation at \(\beta J=0.12\), the non-Gibbs test preparation used above, and a canonical-like distribution treated deliberately as unknown to provide a rare-tail stress test.
Figure~\ref{fig:supp-metrology} compares three calibration goals for each preparation.

For the non-Gibbs mixture used in Table~\ref{tab:supp-arbitrary-states}, the smallest energy population is
\begin{equation}
p_{\min}
=
4.96\times10^{-5}.
\label{eq:supp-pmin-mixture}
\end{equation}
Figure~\ref{fig:supp-metrology} shows that a conservative \(99\%\) all-sectors-seen criterion requires about
\begin{equation}
1.65\times10^5
\label{eq:supp-N-all-seen}
\end{equation}
calibration shots. The local \(1\%\), \(99\%\) task-level estimate from Eq.~\eqref{eq:supp-calibration-var} is of the same order,
\begin{equation}
2.00\times10^5.
\label{eq:supp-N-task-cal}
\end{equation}
By contrast, a conservative uniform requirement \(|\Delta\gamma_n|\leq0.1\) for every sector needs approximately
\begin{equation}
5.91\times10^7
\label{eq:supp-N-uniform-gamma}
\end{equation}
shots.
The canonical-like stress test in Fig.~\ref{fig:supp-metrology} has the very small tail probabilities of the nominal canonical state but treats them as if they were unknown.
Its much larger calibration cost is therefore not a cost that must actually be paid for a known Gibbs preparation.
When the state is known to be canonical, the probabilities are already fixed by \(H_i\) and \(\beta_i\), so no initial-population calibration is needed.

The operational lesson is that the amount of initial information needed depends on the task.
Exact reconstruction of every \(\gamma_n\), control of the fluctuation identity to a chosen tolerance, and full state tomography are different goals and can require very different numbers of measurements.

\end{document}